\documentclass[amsmath,amssymb,aps,eqsecnum,nofootinbib, twocolumn, preprintnumbers]{revtex4}
\usepackage[dvips, dvipdfm]{graphicx, color}
\usepackage{subfigure}
\newcommand{\UV}{U}

\newcommand{\gammatIV}{\tilde\gamma}
\newcommand{\sigmaAH}{\sigma^\text{AH}}

\begin{document}

\title{Time-symmetric initial data of \\
large brane-localized black hole in RS-II model}
\author{%
  Norihiro Tanahashi\footnote{E-mail:tanahashi@tap.scphys.kyoto-u.ac.jp}
  and
  Takahiro Tanaka\footnote{E-mail:tama@scphys.kyoto-u.ac.jp}
  }
\address{%
  Department of Physics, Kyoto University,
  Kyoto 606-8502, Japan\\
}

\preprint{KUNS-2117}

\begin{abstract}
In the aim of shedding a new light on 
the classical black hole evaporation conjecture 
stating that a static brane-localized  black hole (BH) larger than the 
bulk curvature scale does not exist in Randall-Sundrum II
(RS-II) model, we investigate
time-symmetric initial data 
with a brane-localized apparent horizon (AH) and analyzed its properties.
We find that a three-parameter family of such initial data 
can be constructed by simply placing a brane on a constant time 
surface of Schwarzschild anti-de Sitter space. 
By this method, we unambiguously confirm that 
initial data with an arbitrarily large AH area 
do exist.
We compare the ADM mass and the horizon area of our initial data
with that of the black string (BS) solution. If there is a 
sequence of static brane-localized BH solutions, such solutions 
should be contained in the time-symmetric initial data. 
Moreover, if they are stable, it will have a smaller mass
compared with the BS solution with the same horizon area.  
However, we find that any initial data constructed by this method 
do not have a smaller mass than the BS solution 
when the horizon area is larger than the size determined by 
the bulk curvature scale. 
We further investigate what kind of configuration realizes 
the minimum mass for the same AH area. 
The configuration that realizes the smallest mass 
turns out to be the one close to the BS 
truncated by a cap. One may think that this indicates 
the existence of a static brane-localized BH solution.  
However, since our three-parameter family of 
initial data does not include the configuration resembling 
the BS solution, this minimum of mass may just 
reflect the expected minimum of mass corresponding to the 
BS solution. 
We also demonstrate that the same method applies to construct 
initial data in (3+1)-dimensional RS-II brane world. 
In this case an exact solution of a brane-localized BH exists 
but BS solution does not.
Nevertheless, the behavior of the initial data is quite 
similar in both cases. 
We find that the known
exact solution always has a smaller mass 
than our initial data with the same horizon area. 
This result enforces the standard belief that the exact 
BH solution is the most stable black object in the 
four-dimensional RS-II model.
These results are all consistent with 
the classical BH evaporation conjecture, but 
unfortunately it turns out that
they do not provide a strong support of it. 
\end{abstract}
\maketitle

\section{Introduction}
The Randall-Sundrum II (RS-II) model~\cite{Randall:1999vf} 
is a brane world model,
which provides a way to realize our four-dimensional world in a higher-dimensional 
spacetime requested by the string theory or the M-theory.
RS-II model is composed of five-dimensional bulk spacetime 
with negative cosmological constant 
and a four-dimensional brane with positive tension.
Matter fields are confined on the brane while gravity can propagate
through the bulk spacetime.
It is known that 
the weak gravitational field produced by a mass on the brane obeys 
the usual four-dimensional Newton law with a correction 
suppressed at a large distance from the 
source~\cite{Randall:1999vf, Garriga:1999yh, Wiseman:2001xt, Giannakis:2000zx, Kudoh:2001wb},
though the extra-dimension extends infinitely in this model.
This fact means that it is difficult to distinguish this model 
from an ordinary four-dimensional model 
as long as we investigate the weak gravity regime.
Thus we turn focus on the effect of strong gravity.
We study the objects formed after gravitational collapse on the brane, in which 
the effect of strong gravity becomes essential.

Naively, a static black hole (BH)
whose horizon is localized near the brane will be formed 
as a final state of gravitational collapse on the brane.
There is an exact static solution with an event horizon, which is black
string (BS)~\cite{Chamblin:1999by}. 
However, it seems unlikely that a BS is formed as 
a result of gravitational collapse 
since  the BS on the RS-II model is singular and also 
unstable due to so-called Gregory-Laflamme
instability~\cite{Gregory:2000gf}. 
A static solution of a large BH localized on the brane, 
however,  has not been discovered yet, despite lots of effort on 
this issue~\cite{Kanti:2001cj, Karasik:2003tx, Karasik:2004wk,
Dadhich:2000am, Casadio:2001jg, Chamblin:2000ra, 
Shiromizu:2000pg, Kofinas:2002gq, Casadio:2002uv}.
Numerical solution of a static brane-localized BH
has been constructed when the horizon size is not much larger than 
the bulk curvature scale, 
but the construction becomes harder as the horizon size becomes 
larger~\cite{Kudoh:2003xz, Kudoh:2004kf}.
This fact does not exclude the possibility that a static solution of 
brane-localized BH larger than the bulk curvature scale does exist,
but we do not have any strong evidence of its existence.
As an explanation of the lack of static solution, 
there is a conjecture that brane-localized static BHs larger
 than bulk curvature scale do not exist in RS-II model based on the AdS/CFT 
correspondence~\cite{tanaka, emp1}. 

There are several works related to this
conjecture~\cite{Casadio:2004nz, Anderson:2004md,Galfard:2005va,Fabbri:2005zn,Kaloper:2006ek,Fitzpatrick:2006cd,Fabbri:2007kr,Fabbri:2007zk,Tanaka:2007xm},
but no definite conclusion is obtained yet.  
It is desirable to investigate 
the possibility  of the presence
of static black hole solution directly
in order to test the validity of this conjecture.
However, 
it is technically difficult to construct a static
large brane-localized BH numerically. 
Thus, we consider time-symmetric initial data which have a
brane-localized apparent horizon (AH), expecting that 
their properties 
may give some insight into the brane-localized BH.
Notice that all static solutions are contained 
in the time-symmetric initial data.

This paper is organized as follows. 
We will explain the method how we construct 
initial data in Sec.~\ref{Sec2}. 
We find that a three-parameter family of such initial data 
can be constructed by simply placing a brane on a constant time 
surface of Schwarzschild anti-de Sitter (AdS) space. 
In Sec.~\ref{results}, using this method, 
we give an existence proof of initial data with a large AH 
area. We also compare the ADM mass and 
the five-dimensional horizon area of our initial data
with that of the BS solution. 
We further investigate what kind of configuration realizes 
the minimum mass for a given AH area. 
We also compute 
the initial data in (3+1)-dimensional RS brane world, 
in which an exact static solution of brane-localized BH 
does exist~\cite{emp2, Emparan:1999fd}. 
The results are compared with those in (4+1)-dimensional case. 
Section~\ref{Sec fourD} is devoted to summary and conclusion.

\section{Initial data construction method}
\label{Sec2}

In this section we introduce a construction method of 
time-symmetric initial data with a brane-localized AH 
in RS-II model. 
The model is composed of two copies of five-dimensional 
empty bulk with negative cosmological constant $\Lambda$ 
separated by a $\mathbb{Z}_2$-symmetric positive tension brane. 
The tension of the brane satisfies the RS condition
$\lambda=3k/4\pi G_5$ with $k=\sqrt{-\Lambda/6}$, 
where %$\kappa_5^2=8\pi G_5$ and 
$G_5$ is the five-dimensional gravitational constant.
The setup is compatible 
with a Minkowski brane with AdS bulk with the bulk curvature 
length being $k^{-1}$.  
The initial data we consider 
have $O(3)$-symmetry in the spacelike dimension as well as 
the symmetry with respect to time reversal. 
These symmetries are property shared with static brane-localized BH
solutions.
Hence, we think it appropriate to restrict our attention 
to this class of initial data. 

%In this paper we do not newly solve the full set of constraint equations to 
%obtain initial data. 
%Instead, we take a constant time slice from a known vacuum solution 
%with cosmological constant. 
In order to obtain a time-symmetric initial data of the RS-II model, we have to solve 
the Hamiltonian constraint equation in the bulk and on the brane simultaneously.
In this paper, we do that by taking a constant time slice from a known vacuum solution 
with cosmological constant.
We use five-dimensional AdS (topological) BH solutions as such
solutions. 
By using the bulk metric taken from a known solution, 
the constraint equations in the bulk are automatically satisfied. 
In this spacetime we can find spacelike minimal hypersurfaces 
on a constant time slice, on which the
expansion $\Theta$ of 
out-going null geodesic congruence vanishes. Even if those minimal hypersurfaces 
are not closed, they 
are candidates of the AH. 
By cutting this spacetime with a pure-tension brane with 
$\mathbb{Z}_2$-symmetry, a part of AH candidate,  
even if they are not closed originally, can be reformed into a trapped surface. 
In placing a momentarily static brane with $\mathbb{Z}_2$-symmetry, 
we solve the brane configuration starting from a point on a 
$\Theta=0$ hypersurface so as to satisfy the Hamiltonian constraint 
on the brane. 
In this manner we obtain a family of initial 
data with an AH localized on the brane. 
We note here that this construction method is similar to that 
used in Ref.~\cite{Creek:2006je}
to generate a solution of a static brane with brane-localized matter.

\subsection{Bulk solution}

The starting point of our construction procedure is to choose
an asymptotically AdS vacuum solution 
of the Einstein equations with negative cosmological constant $\Lambda$.
In this study, we use the well-known AdS Schwarzschild (the case with $\beta=+1$ below)
solution and its extensions (the cases with $\beta=0$ and $-1$ below),   
which are called ``topological BH in AdS'' \cite{Birmingham:1998nr}. 
The metric is given by 
\begin{eqnarray}
ds^2=-\UV(r)dt^2
%+\UV(r)^{-1}dr^2
+\frac{dr^2}{\UV(r)}
+r^2\sigma_{IJ}(x)dx^Idx^J,
\label{top_metric}
\end{eqnarray}
where 
\begin{eqnarray*}
\UV(r)=\beta+k^2r^2-\frac{\mu}{r^2} \quad  (\beta=+1,0,-1), 
\end{eqnarray*}
and 
\begin{equation*}
\sigma_{IJ}(x)dx^Idx^J=
\begin{cases}
d\chi^2 + \sin^2\chi
%\left(d\theta^2+\sin^2\theta d\phi^2\right)
d\Omega_\text{II}^2
& (\beta=+1)
\\
d\chi^2 + \chi^2
%\left(d\theta^2+\sin^2\theta d\phi^2\right)
d\Omega_\text{II}^2
& (\beta=0) 
\\
d\chi^2 + \sinh^2\chi
%\left(d\theta^2+\sin^2\theta d\phi^2\right)
d\Omega_\text{II}^2
& (\beta=-1) 
\end{cases}
\end{equation*}
are the metrics on the three-dimensional maximally symmetric spaces.
Here $\mu$ is the mass parameter 
and $d\Omega_\text{II}^2=d\theta^2+\sin^2\theta\,d\phi^2$.   
We can see that the spacetime (\ref{top_metric}) is asymptotically AdS, 
since the term $k^2r^2$ in $\UV(r)$ dominates at $r\to\infty$. 

The spacetime described by this metric with $\beta=+1$ has an spherical event
horizon at $r=r_g$ where $\UV(r)$ vanishes. 
Also in the other two cases ($\beta=0$ and $-1$), $r=r_g$ defined by 
$\UV(r_g)=0$ becomes a surface on which the expansion $\Theta$ of 
the outgoing null geodesic congruence vanishes. However, these surfaces are not 
closed unless we assume non-trivial identification in the 
maximally symmetric three space. Nevertheless, we refer to 
the surface defined by $r=r_g$ as an event horizon in all three cases. 
The three spatial metric on the horizon is flat for $\beta=0$  
while it is hyperbolic for $\beta=-1$.
Notice that the event horizon does not 
exist in the case of $\mu< 0$ for $\beta=0, +1$, 
and in the case of $\mu< -1/4k^2$ for $\beta=-1$.
We refer to these solutions as spherical ($\beta=+1$), 
flat ($\beta=0$) and
hyperbolic ($\beta=-1$) AdS BHs, respectively, in this paper.

The AdS BH solutions are connected to each other 
via the flat one. 
Both spherical and hyperbolic AdS BHs reduce to the flat one 
in the limit $\mu\to\infty$.
This can be made transparent by the following coordinate
rescaling:
\begin{equation}
\bar{t}=\left(k^2\mu\right)^{1/4}t,\quad 
\bar{r}=\left(k^2\mu\right)^{-1/4}r,\quad
\bar{\chi}=\left(k^2\mu\right)^{1/4}\chi.
\label{rescaling}
\end{equation}
Then, the metric for $\beta=\pm 1$ in the limit 
$\mu\to\infty$ becomes
\begin{align}
ds^2\xrightarrow{\mu\to\infty}
&-\left(k^2\bar{r}^2-\frac{1}{k^2\bar{r}^2}\right)d\bar{t}^2
+\left(k^2\bar{r}^2-\frac{1}{k^2\bar{r}^2}\right)^{-1}
\!\!\!\!d\bar{r}^2
\notag \\
&\qquad\qquad\qquad\;\;
+\bar{r}^2\left(
d\bar{\chi}^2+\bar{\chi}^2d\Omega_\text{II}
\right).
\label{rescaled eq}
\end{align}
This is nothing but the metric for $\beta=0$ after the 
rescaling (\ref{rescaling}).  
In this expression $\mu$ is not present any more. 
This means that the flat AdS BH has only one free parameter $k$.

In the following discussion we set $k$ to unity by rescaling 
the unit of length. In this sense, this background spacetime 
has only one free parameter $\mu$, which 
becomes one of free parameters of the initial data we construct
below.

\subsection{AH candidates in AdS BH spacetime}

Let us consider a three surface 
on a $t=$constant hypersurface $\Sigma_t$. 
If the expansion $\Theta$
of the outgoing null geodesic congruence
on this three-surface vanishes, it 
becomes a candidate of an AH.
Even if the surface is not closed in the original spacetime, 
it might be made compact after we introduce 
a $\mathbb{Z}_2$-symmetric brane. 
Hence, we refer to such a surface with $\Theta=0$ as an apparent horizon 
candidate (AHC).

We denote the unit vector normal to an AHC 
in $\Sigma_t$ as $s^i$. Here Latin indices 
starting from the middle of the alphabet $(i,j,\cdots)$ 
run over all spatial coordinates. 
Then the condition of vanishing expansion of 
the outgoing null geodesic congruence
emanating from this AHC 
is given by 
\begin{equation}
    K-K_{ij}s^i s^j-D_i s^i=0,
\end{equation}
where $K_{ij}$ is the extrinsic curvature of the surface $\Sigma_t$
and $K$ is its trace. 
$D_i$ is the covariant differentiation with respect to 
the induced metric on $\Sigma_t$. 
Since we have $K_{ij}=0$ by the assumption of time-symmetric initial data, 
this equation is reduced to 
\begin{equation}
    D_i s^i=0, 
\label{eqAH}
\end{equation}
which determines the position of the AHC.

Assuming $O(3)$-symmetry of AHC,  
we specify its trajectory by $(r, \chi)=(r_\text{AH}(\zeta), \chi_\text{AH}(\zeta))$,
where $\zeta$ is the proper radial length along the AHC
measured from the axis of the $O(3)$-symmetry.
These $r_\text{AH}(\zeta)$ and $\chi_\text{AH}(\zeta)$ satisfy
\begin{equation}
U^{-1}r'^2_\text{AH}+ r_\text{AH}^2\,\chi'^2_\text{AH}=1,
\label{constraint}
\end{equation}
where a prime means a differentiation with respect to the argument,
which is $\zeta$ here. 
Then, the spacelike unit vector normal to the AHC is 
\begin{equation}
s_\mu=\frac{r_\text{AH}}{\sqrt{U}}
\left(\chi'_\text{AH}, -r'_\text{AH}  \right).
\label{branetangents_5D}
\end{equation}
Then Eq.~(\ref{eqAH}) and Eq.~(\ref{constraint}) can be recasted into a set of two ordinary 
differential equations, whose  explicit form for the spherical AdS BH bulk is given by
\begin{align}
 \frac{\sqrt{U}r_\text{AH}}{r'_\text{AH}}\chi''_\text{AH}
+4
&
\sqrt{U}\chi'_\text{AH}
-\frac{2\cot\chi_\text{AH}}{\sqrt{U}r_\text{AH}}r'_\text{AH}
=0,
\notag \\
-\frac{1}{\sqrt{U}r_\text{AH}\chi'_\text{AH}} r''_\text{AH}
+&
3\sqrt{U}\chi'_\text{AH}
\notag \\
+\frac{r'_\text{AH}}{2U^{3/2}r_\text{AH}\chi'_\text{AH}}
&\left(
\frac{dU}{dr}r'_\text{AH}  %U'
-4U\chi'_\text{AH}\cot\chi_\text{AH}\right)
=0.
\label{eqAH_}
\end{align} 
The expression for the hyperbolic AdS BH bulk is 
obtained by simply replacing $\cot\chi_\text{AH}$ with 
$\coth\chi_\text{AH}$.

We solve this equation setting $\chi_\text{AH}(0)=0$. 
We can freely choose the value of $r_\text{AH}(0)$, which 
specifies the position of the AHC in the background spacetime. 
This $r_0^\text{AH}\equiv r_\text{AH}(0)$ 
becomes one of free parameters of the initial data we construct.
The boundary condition  at $\zeta=0$ is given by 
\begin{equation}
    r'_\text{AH}(0)=0, 
\end{equation}
which comes from the regularity of the AHC on the axis. 
We solve Eq.~(\ref{eqAH_}) with this boundary condition
numerically to obtain the trajectories of AHCs. 

\subsection{Brane trajectory}

Next we put a vacuum brane with $\mathbb{Z}_2$-symmetry   
in the AdS BH bulk. 
We denote the unit normal 
of the brane by $\tilde s_\mu$. 
We take this $\tilde{s}_\mu$ in the direction 
toward the bulk from the brane. 
We introduce the induced metric 
$\gammatIV_{\mu\nu}\equiv g_{\mu\nu}-\tilde{s}_\mu\tilde{s}_\nu$ on the brane. 
The extrinsic curvature $\tilde{K}_{ab}$ on the brane is defined by 
$\tilde{K}_{ab}=-\gammatIV_a^{\;\;\mu} 
\gammatIV_b^{\;\;\nu}\nabla_\mu\tilde{s}_\nu$.
Here Latin indices starting from the beginning of the alphabet $(a, b,
\cdots)$ run the four-dimensional coordinates on the brane.

A vacuum brane has 
the four-dimensional energy-momentum tensor localized on the brane given by 
\begin{equation}
    T_{ab}=-\lambda\gammatIV_{ab}. 
\end{equation}
Israel's junction condition~\cite{Israel:1966rt} on the brane is given by 
\begin{equation}
    \tilde{K}_{ab}-\tilde{K}\gammatIV_{ab}
%=\frac{1}{2}\kappa_5^2 T_{ab}, 
=\frac{1}{2}\cdot 8\pi G_5 T_{ab}, 
\label{junction}
\end{equation}
where we used $\mathbb{Z}_2$-symmetry across the brane. 
This junction condition is just a part of the Einstein equations 
integrated across the brane. 
The equations for the initial data to satisfy is the constraint 
equations, i.e., the ($t,t$)- 
and ($t,i$)-components of the Einstein equations.  
They are the Hamiltonian constraint and momentum constraints respectively.
Since the momentum constraints are trivially satisfied
at the moment of the time-reversal symmetry,  
we only have to consider the Hamiltonian constraint, 
which is ($t,t$)-component of Eq.~(\ref{junction}). 
Using the normal vector $\tilde s^\mu$, this condition is written as 
\begin{equation}
   D_i \tilde s^i=-3k. 
\label{HC}
\end{equation}

As before, we assume $O(3)$-symmetry, and
we specify the brane trajectory by $(r,\chi)=(r_b(\xi), \chi_b(\xi))$, 
where $\xi$ is the proper radial length along the brane. 
The spacelike unit normal $\tilde{s}_\mu$ is given by Eq.~(\ref{branetangents_5D}),
replacing $r_\text{AH}$ and $\chi_\text{AH}$ with $r_b$ and $\chi_b$ respectively.
Then the Hamiltonian constraint~(\ref{HC}) becomes a second order ODE
of $r_b(\xi)$ and $\chi_b(\xi)$. The explicit expression for the spherical AdS BH becomes 
Eq.~(\ref{eqAH_}), replacing $r_\text{AH}$ and $\chi_\text{AH}$ with $r_b$ and $\chi_b$
on the left hand sides, and $0$ with $-3k$ on the right hand sides.

As we are not interested in the spacetime interior of the AHC, 
we solve the brane trajectory from a point on the AHC. 
The choice of the starting point on the AHC is arbitrary. 
This degree of freedom becomes one of free parameters of the initial data we construct.
The boundary condition for Eq.~(\ref{HC}) at this point is 
determined by the regularity of the AH across the brane. 
Namely, the AH should intersect the brane perpendicularly, 
i.e. $\tilde{s}_\mu s^\mu=0$.
This leads to the condition, 
\begin{equation}
\left(r'_b, \chi'_b \right)
=
\left(
\sqrt{U}r_\text{AH}\,\chi'_\text{AH}, -\frac{r'_\text{AH}}{\sqrt{U}r_\text{AH}}
\right),
\label{braneBC}
\end{equation}
at the crossing point.
We solve Eq.~(\ref{HC}) numerically to obtain the brane trajectory. 
Once the AHC is truncated by the brane, it becomes a closed surface with 
vanishing expansion, $\Theta=0$. 
However, the AH is not simply a compact surface with $\Theta=0$ but 
it must be the outermost one among such surfaces. 
Thus we have to check if there is no other $\Theta=0$ hypersurface 
in the region outside of the AHC. 
If such surfaces exist, the outermost one is the true AH. 
This true AH search can be done by the shooting method. 
We take a point on the axis at $\chi=0$, 
and we extend the hypersurface with $\Theta=0$ from the point 
perpendicularly to the axis. 
This hypersurface may hit the brane at an angle. 
Moving the starting point on the axis, we search for the case 
that the hypersurface with $\Theta=0$ crosses the brane perpendicularly. 
If there is no such a hypersurface with $\Theta=0$, 
the original AHC is the genuine AH.  
In the case that true AH exists outside AHC, such initial data will 
be also given by other values of free parameters in the parameter space.
Hence, we just discard such initial data, and analyze only the data
which have no outer AH.

In the case that no event horizon exists in the original bulk solution, 
the original spacetime has a naked curvature singularity at $r=0$,
at which the Kretschmann scalar diverges as $\mathcal{O}(\mu^2/r^{8})$.
However, even in this case, AH may be formed after a brane is 
placed.
Then, this singularity cannot be seen from an observer outside the 
horizon if the brane and the AH hide the singularity. 
In this case there is no problem in adopting this singular bulk 
to construct initial data. 
Nevertheless, we do not consider this case in the following discussion
simply because such initial data do not show any interesting properties 
as a matter of fact.

\begin{figure}[htbp]
\centering
\includegraphics[width=5cm, clip]{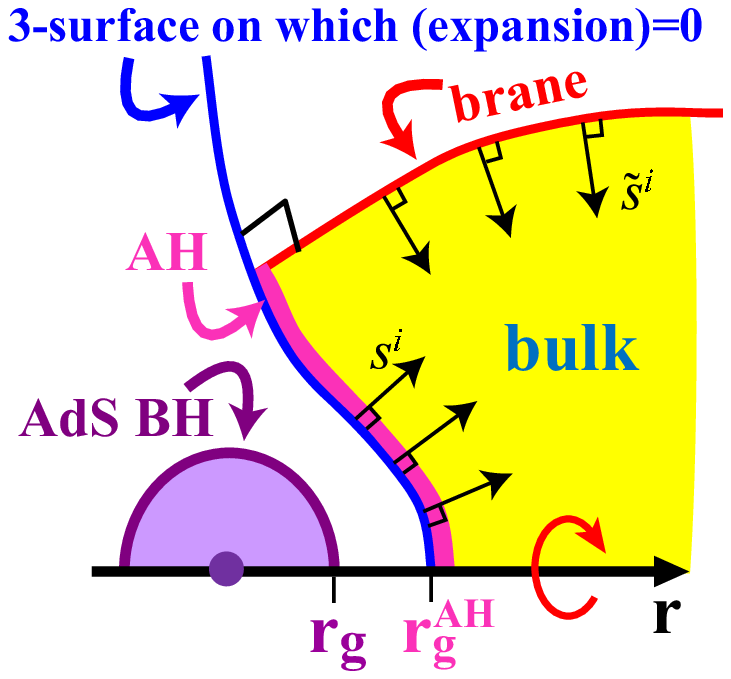}
\caption{A schematic picture of the initial data that we construct.
The whole initial data is composed by gluing two copies of 
the portion presented in this figure together along the brane. 
This initial data has three degrees of freedom: the mass parameter $\mu$ of the 
AdS BH, the position of the AH on the axis $r=r_0^\text{AH}$ and  
the position of the brane starting point. 
For convenience, we take the AH area, $A_\text{AH}$, 
in place of the last parameter. 
}
\label{illust}
\end{figure}

\subsection{ADM mass and AH area}
To characterize the initial data constructed by the method 
explained above, we discuss how to extract 
the ADM mass measured on the brane and the AH area. 

The ADM mass measured on the brane is calculated 
from the spatial part of the 
induced metric on the brane, 
\begin{equation}
    d\tilde{l}^2
= 
%\left\{\UV^{-1}+\left(\frac{d\chi_b}{dr}\right)^2\right\}
%\left(\frac{dr}{d\rho}\right)^2 d\rho^2
\left(\frac{d\rho_b}{d\xi}\right)^{-2} \!\!d\rho_b^2
%	+\rho_b^2 (d\theta^2+\sin^2\theta d\phi^2),
	+\rho_b^2 d\Omega_\text{II}^2,
\end{equation}
where we introduced a new radial coordinate 
$\rho_b(\xi)\equiv r_b(\xi)\sin\chi_b(\xi)$ 
for the spherical AdS BH background case,
corresponding to the circumferential radius of 
$(\theta,\phi)$-two sphere. 
For the hyperbolic AdS BH case, $\sin\chi_b$ in $\rho_b$ is replaced with $\sinh\chi_b$.

Asymptotically, 
$\rho'_b$ behaves as
\begin{equation}
 \rho'_b\approx 1-\frac{\tilde{G}_4 M_\text{ADM}}{\xi} + \cdots,
\end{equation}
where $M_\text{ADM}$ is the four-dimensional ADM mass and
$\tilde{G}_4=kG_5$ is the effective four-dimensional Newton constant. 
%the solution of 
%%(\ref{HC_}) 
%{\color{red} (\ref{HC})} 
%behaves as 
%\begin{equation}
%\chi_b(r)\approx {c_{1/2}\over\sqrt{r}}+{c_{1}\over r}+\cdots,  
%\label{asymchi}
%\end{equation}
%with $c_i$ being constants. The equation 
%(\ref{HC_}) 
%{\color{red} (\ref{HC})} 
%gives relation among the coefficients $c_i$ but first two, 
%$c_{1/2}$ and $c_1$, are not constrained.  
%Substituting the asymptotic form (\ref{asymchi}) into Eq.~(\ref{ADM mass}), 
%we find that the ADM mass is given by 
%%$M_\text{ADM}=16\pi\kappa_5^{-2} c_1$. 
%{\color{blue}$M_\text{ADM}=16\pi c_1 / 8\pi G_5$. }
%In order to 
%determine the coefficient $c_1$ accurately, 
%$\chi_b(r)$ must be solved up to a large enough $r$.
%In the present method, which only requires 
%to solve the constraint equation along the brane trajectory, 
%we do not have to solve the 
%Hamiltonian constraint equation in the bulk. 
%This enables us to determine $\chi_b(r)$ up to very large $r$.
%This is an advantage of this method.  
%{\color{red}
%We note here that the coefficient $C_{1/2}$ is nothing but 
%the parameter which labels the position of the brane
%in the asymptotic region $r\to\infty$.
In order to determine $M_\text{ADM}$ accurately, 
$r_b(\xi)$ and $\chi_b(\xi)$ must be solved up to a large enough $\xi$.
In the present method, which only requires 
to solve the constraint equation along the brane trajectory, 
we do not have to solve the 
Hamiltonian constraint equation in the bulk. 
This enables us to determine $\rho_b(\xi)$ up to very large $\xi$.
This is an advantage of this method.

The AH area is also an important physical quantity, which plays 
an important role in the following discussion. 
The spatial part of the induced metric on the AH is
\begin{equation}
\sigmaAH_{IJ}dx^Idx^J=
d\zeta^2
+
%r_\text{AH}^2\sin^2\chi \;
\rho_\text{AH}(\zeta)^2
%\left(d\theta^2+\sin^2\theta\, d\phi^2\right)
d\Omega_\text{II}^2
\end{equation}
in the spherical AdS BH case, where 
$\rho_\text{AH}(\zeta)\equiv r_\text{AH}(\zeta)\sin\chi_\text{AH}(\zeta)$.
For the hyperbolic AdS BH background case, 
$\sin\chi_\text{AH}$ in this expression is replaced with $\sinh\chi_\text{AH}$.
Then the area of the AH, $A_\text{AH}$, is calculated as 
\begin{equation}
A_\text{AH}
= \int \sqrt{\sigmaAH} d\chi d\theta d\varphi 
=
2 \int_0^{\zeta_0} 
4\pi\, 
%r_\text{AH}^2\sin^2\chi_\text{AH} \; 
\rho_\text{AH}{}^2 \,d\zeta,
\label{area_5D}
\end{equation}
where $\zeta_0$ is the value of $\zeta$ 
at which the brane crosses the AH.

\section{Analysis of the properties of initial data}
\label{results}

\subsection{Existence of initial data with a large AH}

The family of initial data constructed by the method explained above 
has three free parameters: the mass parameter of the AdS BH bulk $\mu$, 
the position of the starting point of AH $r_0^\text{AH}$ 
and the location of the brane starting point. 
We will use the area of the AH, $A_\text{AH}$, instead of the last parameter.
We do not count the bulk curvature length scale $k^{-1}$ 
as a parameter since we can absorb it 
by rescaling the unit of length. 
Since there is no static brane-localized BH solution with 
a large horizon area, one may suspect that 
time-symmetric initial data with a large AH do not exist. 
However, we found no difficulty in constructing initial data 
with a large AH.
 
As we mentioned above, the original AHC may not 
be a true AH if another AH is found outside of it. 
%When we fix the AH area, we still have two remaining free parameters, $\mu$ and $r^\text{AH}_0$. 
When we fix one of three free parameters, we still have two remaining free parameters.
In Fig.~\ref{long.eps} and \ref{contour5D}, the parameter regions in which another AH is 
found outside of the original AHC are shaded, and 
those in which the brane is not asymptotically flat
are dotted. 
%Different plots correspond to different values of the fixed AH area. 
%The left (right) half of each plot corresponds to the initial data 
%which have the spherical (hyperbolic) AdS BH bulk. 
We refer to the shaded and dotted regions as the excluded regions. 
The excluded region is 
confined to the region where $r^\text{AH}_0$ is close 
to the BH horizon of the original AdS BH irrespective of the size of 
$A_\text{AH}$.

\subsection{Area comparison with RS-II black string}
\label{sec:BS}
The only known exact solution containing a black object in the RS-II model 
is the BS solution. 
The BS solution in the RS-II model is given by 
\begin{equation}
    ds^2=\frac{1}{k^2z^2}\left( 
-f(r) dt^2
%+ f(r)^{-1}dr^2 +r^2 (d\theta^2+\sin^2\theta\,d\phi^2)+dz^2
+ f(r)^{-1}dr^2 +r^2 d\Omega_\text{II}^2 +dz^2
\right),
\label{metric_BS}
\end{equation}
where
$f(r)=1-r_g/r$ and $r_g$ is the horizon radius on the
brane. $r_g$ is related to the ADM mass measured on the brane as
$M_\text{BS}=r_g/2\tilde G_4$.
The brane is at $z=k^{-1}$ in these coordinates.
The area of the BS horizon $A_\text{BS}$ is 
\begin{equation}
    A_\text{BS} = 2\int_{k^{-1}}^\infty  
  \frac{4\pi r_g^2}{(kz)^3} \; dz
= 4\pi r_g^2 k^{-1}.
\end{equation}
We compare this area with the AH area $A_\text{AH}$ in the initial data,
equating 
%the ADM mass 
$M_\text{ADM}$ of the initial data
%(\ref{ADM mass})
%to that of BS 
to $M_\text{BS}$.

What we want to know is whether one can construct initial data 
with a larger horizon area than BS when 
the ADM mass is fixed. Why is it an interesting question? 
If there is a static localized BH solution, it 
gives a time-symmetric initial data. 
Suppose that there is 
a static localized BH solution. Then, if we can survey 
all possible time-symmetric initial data, the static solution must 
be found among them. Moreover, if the solution is stable, we expect that 
it will realize the maximum horizon area among all configurations 
with the same mass. Hence, its horizon area must be greater than 
that of BS, too.  
Hence, if we find a time-symmetric initial data 
that has the horizon area larger than BS, 
it suggests that there is a static brane-localized BH solution\footnote{ 
Of course, our statement here is naive. Even if there are 
initial data with the horizon area larger than BS, 
we cannot claim that it proves the existence of a static 
brane-localized BH solution. 
One possible loophole is that 
such initial data may not evolve into a stationary state. 
One may point out that the configuration that realizes the maximum 
of the horizon area for a fixed ADM mass will exist. Such a
configuration seems to be static. This statement might be true, 
but we would not be able to deny the possibility that 
such a configuration might be singular. In fact, if all
time-symmetric initial data have a smaller horizon area than 
that of the BS solution, the maximum of the horizon area 
is realized by the BS solution, which is singular.  
All the initial data close to BS will be unstable 
due to the Gregory-Laflamme instability, 
and they will not settle into a regular stationary state. 
}.

\begin{widetext}

\begin{figure}[htbp]
\centering
\subfigure[Spherical AdS BH background, $\mu<\mathcal{O}(k^{-2})$ ]
{\includegraphics[width=6.4cm, clip]{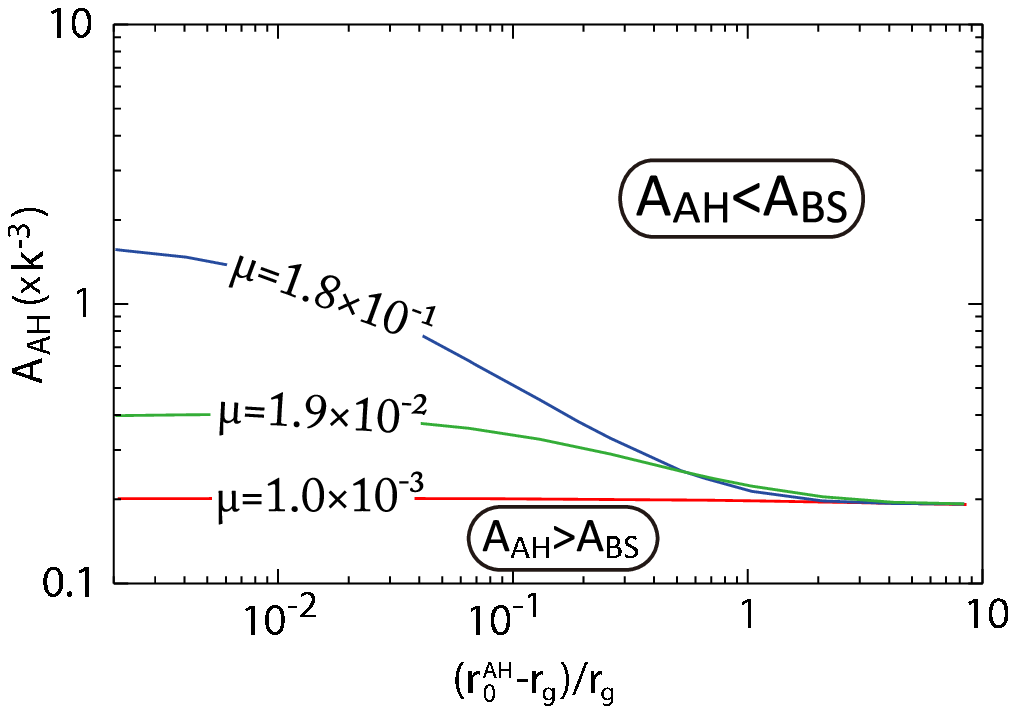}
}
\subfigure[Spherical AdS BH background, $\mu>\mathcal{O}(k^{-2})$]
{\includegraphics[width=6.4cm, clip]{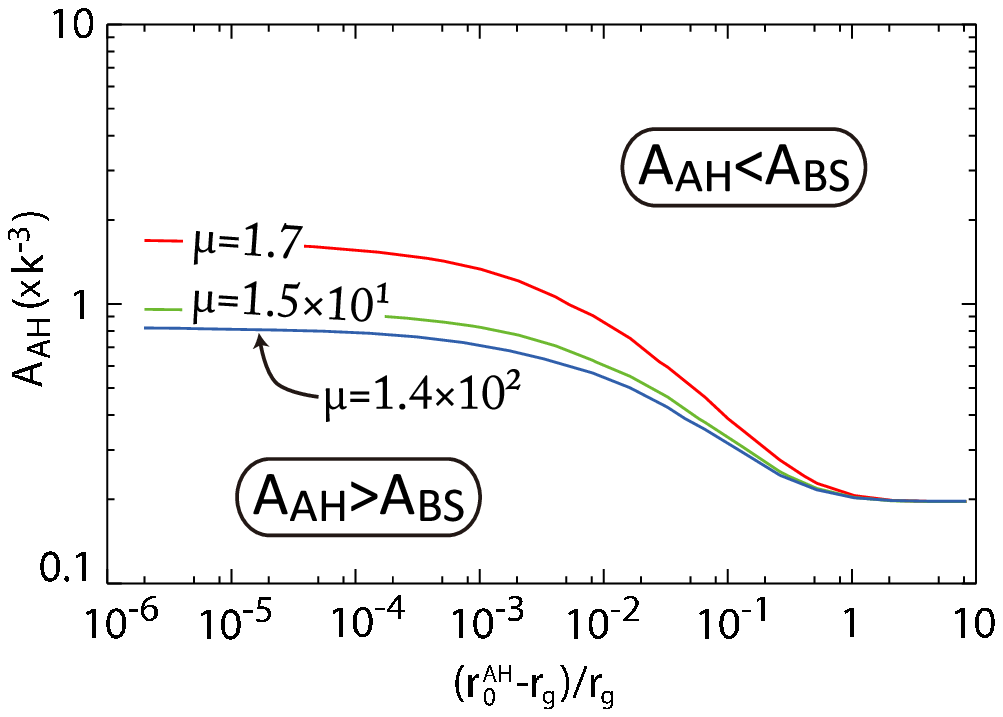}
}
\subfigure[Hyperbolic AdS BH background, $\mu=(-1/4+10^{-2})k^{-2}$]
{\includegraphics[width=6.4cm, clip]{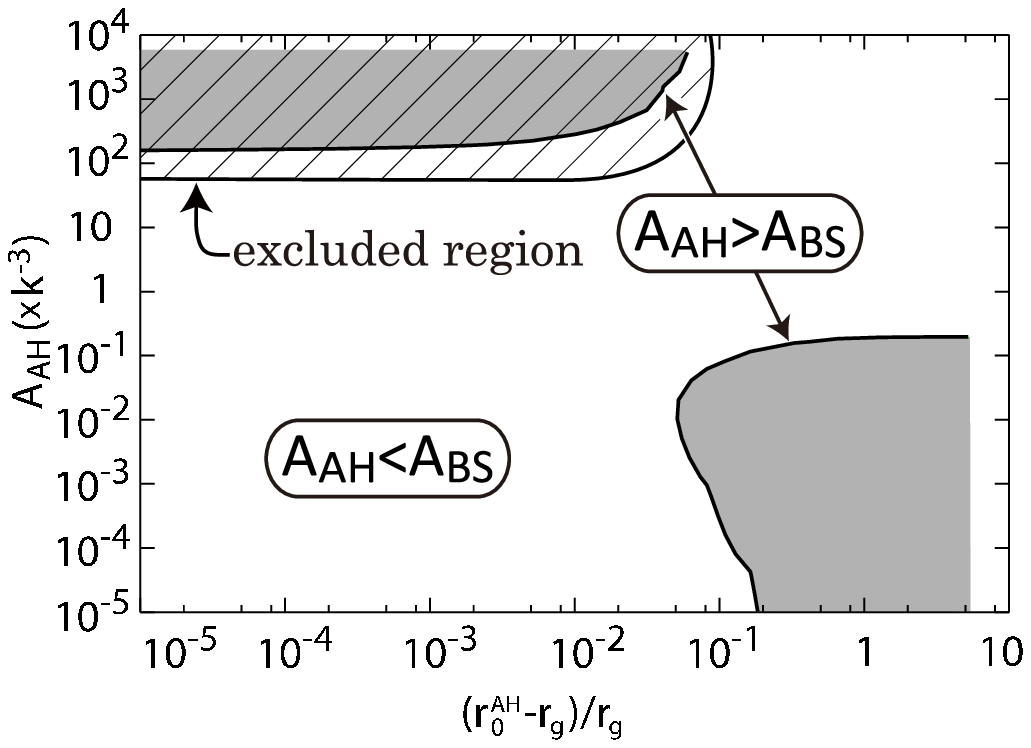}
}
\subfigure[Hyperbolic AdS BH background, $\mu=10k^{-2}$]
{\includegraphics[width=6.4cm, clip]{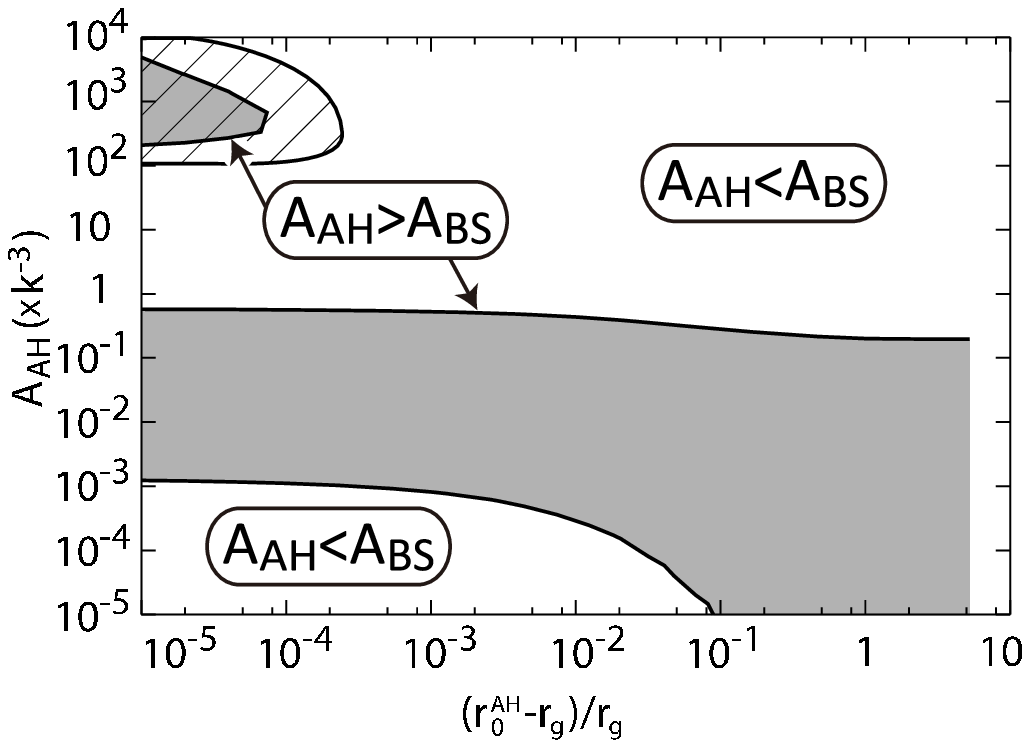}
}
\caption{
The boundaries between two cases, $A_\text{AH}<A_\text{BS}$ and $A_\text{AH}>A_\text{BS}$,  
are shown for several values of $\mu$. 
The horizontal axis $(r_0^\text{AH}-r_g)/r_g$, which is 
the distance between the AH and the event horizon of AdS BH 
normalized by $r_g$. 
The vertical axis is the AH area.
The AH area is larger (smaller) than the BS horizon area below (above) 
the boundary curve. 
The panels (a) and (b) are the figures in the case of the spherical AdS BH background
for $\mu<\mathcal{O}(k^{-2})$ and $\mu>\mathcal{O}(k^{-2})$ respectively.
 The region $(r_0^\text{AH}-r_g)/r_g<10^{-3}$ is not shown in the panel (a),
because the boundary is mostly inside the excluded region. 
The panels (c) and (d) are the figures in the case of the hyperbolic AdS BH background
for $\mu=(-1/4+10^{-2})k^2$ and $\mu=10k^{-2}$.
In the gray region the AH area is larger
than the BS horizon area.
The initial data in the shaded region have an outer AH and they must be excluded from the data set
to conduct the horizon area comparison.}
\label{long.eps}
\end{figure}

\end{widetext}

Now we show the results of the comparison between $A_\text{AH}$ and 
$A_\text{BS}$. In Fig.~\ref{long.eps}, 
we show the boundaries in the parameter space 
between two cases, $A_\text{AH}<A_\text{BS}$ and $A_\text{AH}>A_\text{BS}$. 
For the purpose of display, we reduced the parameter space to 
two dimensions by fixing the value of $\mu$.  
The boundary curves are drawn in the parameter space 
$\left((r_0^\text{AH}-r_g)/r_g, A_\text{AH}\right)$ 
for several values of $\mu$.

The panel (a) shows the border lines of the two regions for 
$\mu=1.8 \times 10^{-1}k^{-2}$, $1.9\times 10^{-2}k^{-2}$ and $1.0\times 10^{-3}k^{-2}$, 
while the panel (b) for 
$\mu=1.7k^{-2}$, $1.5\times 10^1k^{-2}$ and $1.4\times 10^2k^{-2}$ 
in the case of the spherical AdS BH background.
In the panel (a)
 the region $(r_0^\text{AH}-r_g)/r_g<10^{-3}$
is not shown since the boundary is mostly inside the excluded region. 
We found that the boundaries are at around 
$A_\text{AH} \sim \mathcal{O}(k^{-3})$ irrespective of 
the value of $\mu$ and $\left(r_0^\text{AH}-r_g\right)/r_g$.
The panels (c) and (d) are the figures in the case of the hyperbolic AdS BH background
for $\mu=(-1/4+10^{-2})k^{-2}$ and $\mu=10k^{-2}$.
In the gray region the AH area is larger
than the BS horizon area.
The initial data in the shaded region have an outer AH and they are excluded from the data set
to conduct the horizon area comparison. 
We found that the AH area never becomes larger than the BS area
when $A_\text{AH}\gtrsim\mathcal{O}(k^{-3})$ even in this case.
Namely, all the initial data that we can construct by the present method 
have a horizon area smaller than that of BS with the same ADM mass 
if the AH area is larger than a typical size determined by the bulk
curvature scale. 
This result is consistent with the classical BH evaporation conjecture, 
though it does not exclude the existence of counter-examples. 

\subsection{Initial data with maximum horizon area for a fixed mass}
\label{Sec contour5D}
As we stated above, if there is a static brane-localized BH solution, 
it must be contained in time-symmetric initial data, if we take into
account all possible configurations. Moreover, if the solution is
stable, we expect that it will be realized as a maximum of the horizon area among all
time-symmetric initial data with the same ADM mass. 
Although we can search only a limited part of time-symmetric 
initial data by our method, it will be still interesting to 
see which configuration realizes the maximum horizon area 
when the ADM mass is fixed. 

Technically, it is easier to fix the AH area than 
to fix the ADM mass, because $A_\text{AH}$ 
itself is one of the three parameters that label our initial data. 
To the contrary, in order to calculate $M_\text{ADM}$, 
the information in the asymptotic region $r\to\infty$ is needed. 
The initial data that realizes 
the maximum horizon area with the ADM mass fixed 
should have the minimum ADM mass when the AH area is fixed. 
Therefore we fix the AH area and search for 
the initial data of 
the minimum ADM mass varying the other parameters, 
$\mu$ and $r^\text{AH}_0$.

\begin{widetext}

 \begin{figure}[!h]
\centering
\subfigure[$A_\text{AH}=k^{-3}$, ($M_\text{BS}\simeq 0.141k^{-1}G_5^{-1}$)]
{\includegraphics[width=8.3cm, clip]{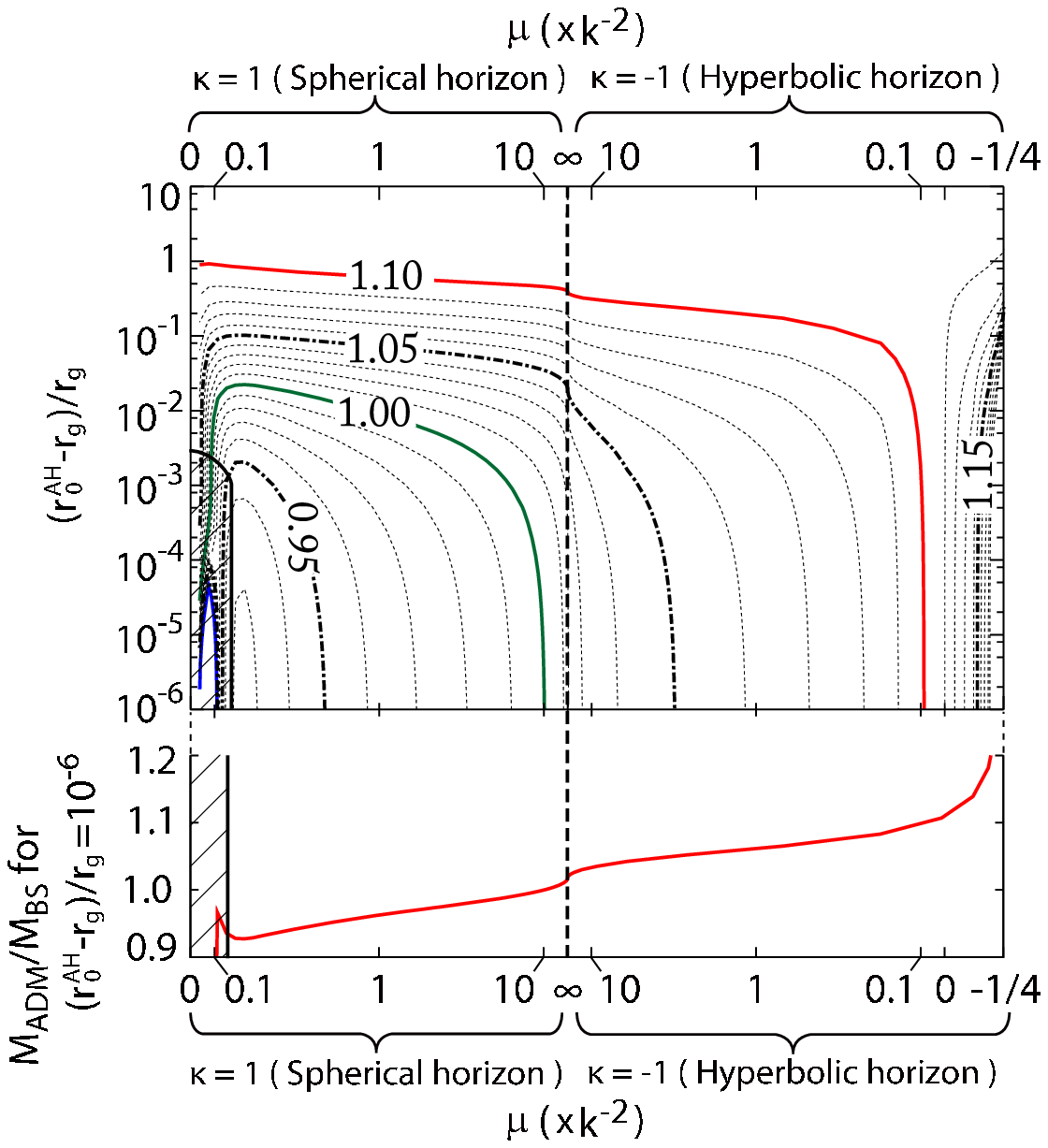}
}
\subfigure[$A_\text{AH}=10k^{-3}$, ($M_\text{BS}\simeq 0.446k^{-1}G_5^{-1}$)
]
{\includegraphics[width=8.3cm, clip]{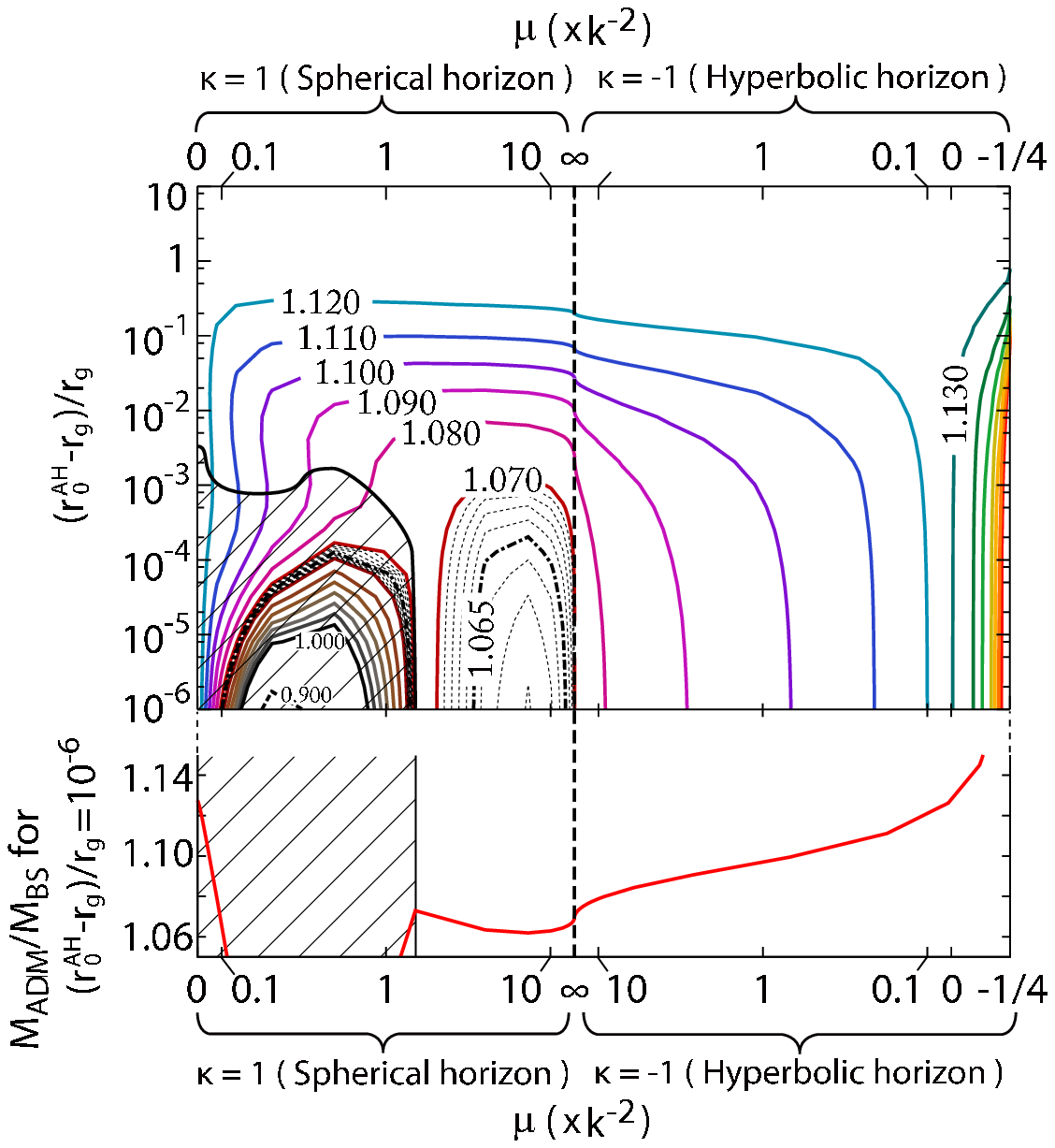}
}
\subfigure[$A_\text{AH}=10^{2}k^{-3}$ ($M_\text{BS}\simeq 1.41k^{-1}G_5^{-1}$)]
{\includegraphics[width=8.3cm, clip]{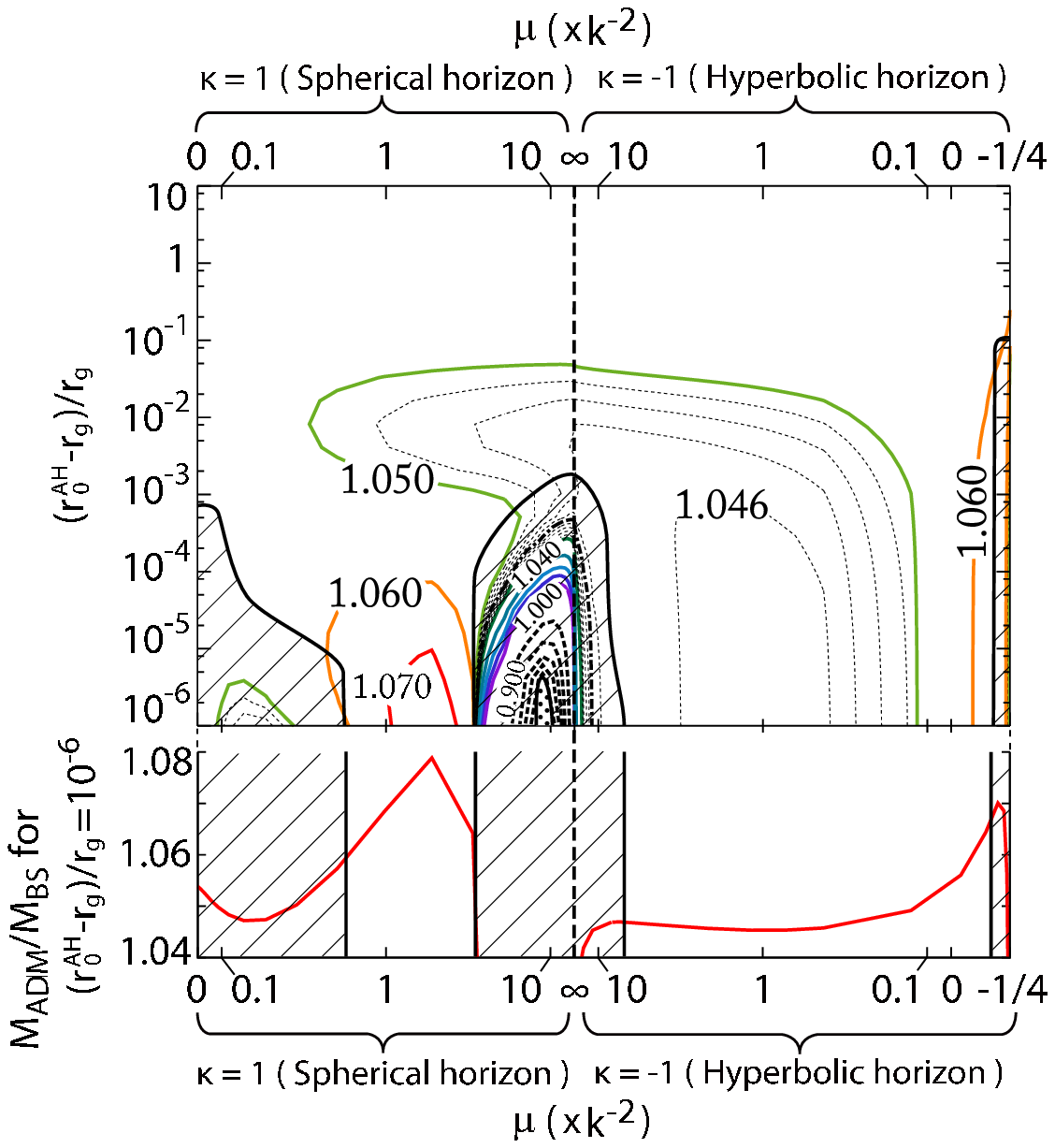}
}
\subfigure[$A_\text{AH}=10^{3}k^{-3}$ ($M_\text{BS}\simeq 4.46k^{-1}G_5^{-1}$)]
{\includegraphics[width=8.3cm, clip]{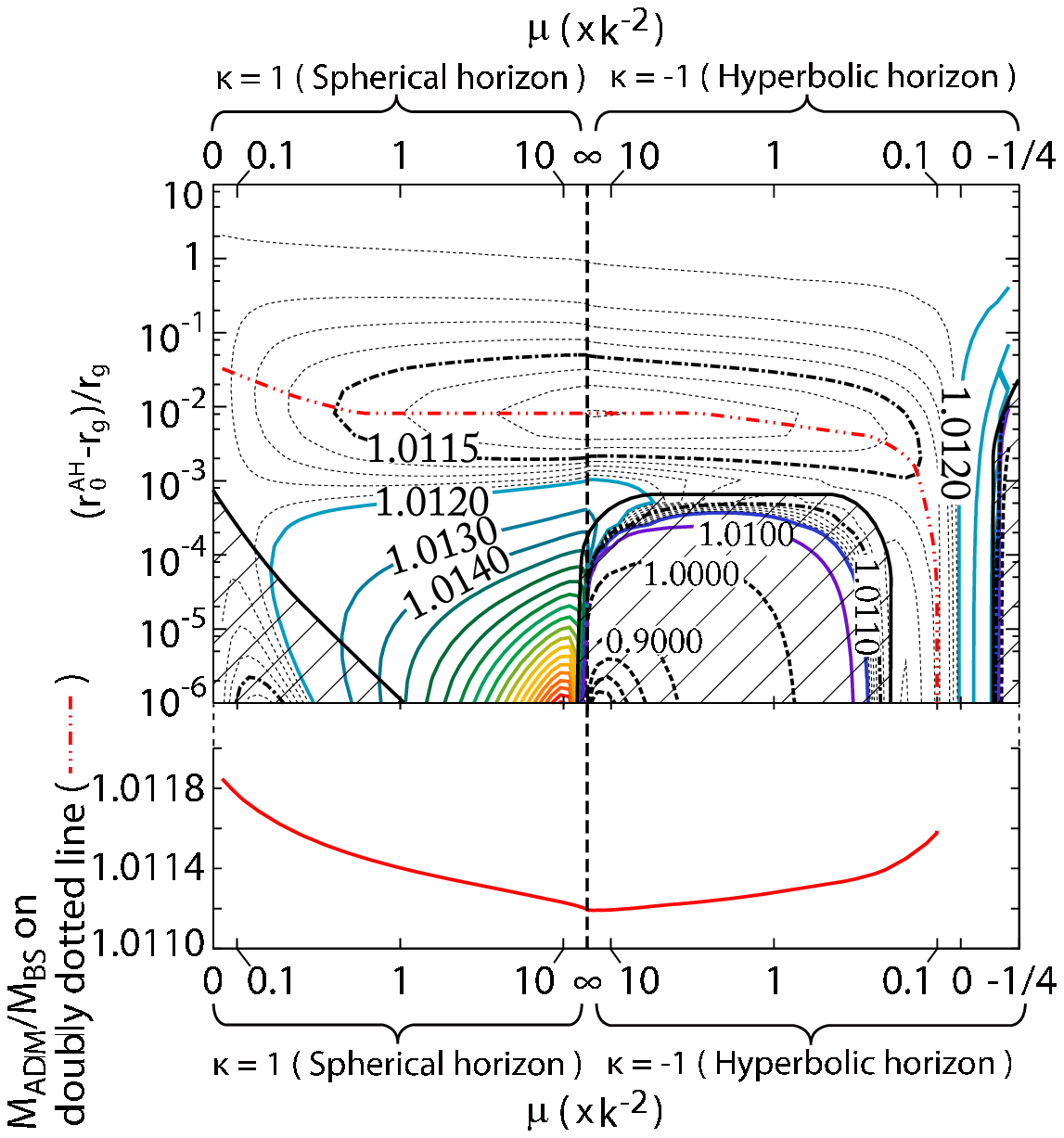}
}
\caption{
The contour plots of $M_\text{ADM}/M_\text{BS}$ 
on the parameter space $\left(\mu, (r^\text{AH}_0-r_g)/r_g\right)$
for several values of $A_\text{AH}$,
where $M_\text{BS}$ is the ADM mass of BS whose horizon area
is equal to $A_\text{AH}$.
The horizontal axis is the value of $\mu$ for the spherical 
and hyperbolic AdS BH cases. 
The curves below the contour plots show the values of $M_\text{ADM}$ for
$(r_0^\text{AH}-r_g)/r_g=10^{-6}$ in the panels (a), (b) and (c), 
and the value along the doubly dotted curve drawn 
in the contour plot in the panel (d). 
The initial data in the shaded region have an outer AH and they are
excluded from the search for the mass minimum.
For the parameters in the dotted region in 
the panel (c), 
the initial data do not have an asymptotically flat brane, so they are
also excluded from the data set.
The mass minimum is realized on the spherical AdS BH background 
in the panel (a) and (b), 
%\ref{A1e0.eps} and \ref{A1e1.eps},
on the hyperbolic AdS BH background in the panel (c),
and on the flat AdS BH background in the panel (d). 
}
\label{contour5D}
\end{figure}

\end{widetext}

We have drawn contour plots of 
$M_\text{ADM}/M_\text{BS}$for several values of $A_\text{AH}$ in Fig.~\ref{contour5D}, 
where $M_\text{BS}$ is the ADM mass of a BS whose horizon area is equal to $A_\text{AH}$.  
Different plots correspond to different values of the fixed AH area. 
The left (right) half of each plot corresponds to the initial data 
which have the spherical (hyperbolic) AdS BH bulk. 
Here we should remember that the shaded and dotted regions should be neglected
because there is a true AH outside the original AHC. 
Once we exclude these regions, 
we find that there is one and only one minimum of $M_\text{ADM}$ in 
all contour plots.

\begin{widetext}
 
\begin{figure}[htbp]
\centering
\subfigure[$A_\text{AH}=k^{-3}$]
{\includegraphics[width=5cm, clip]{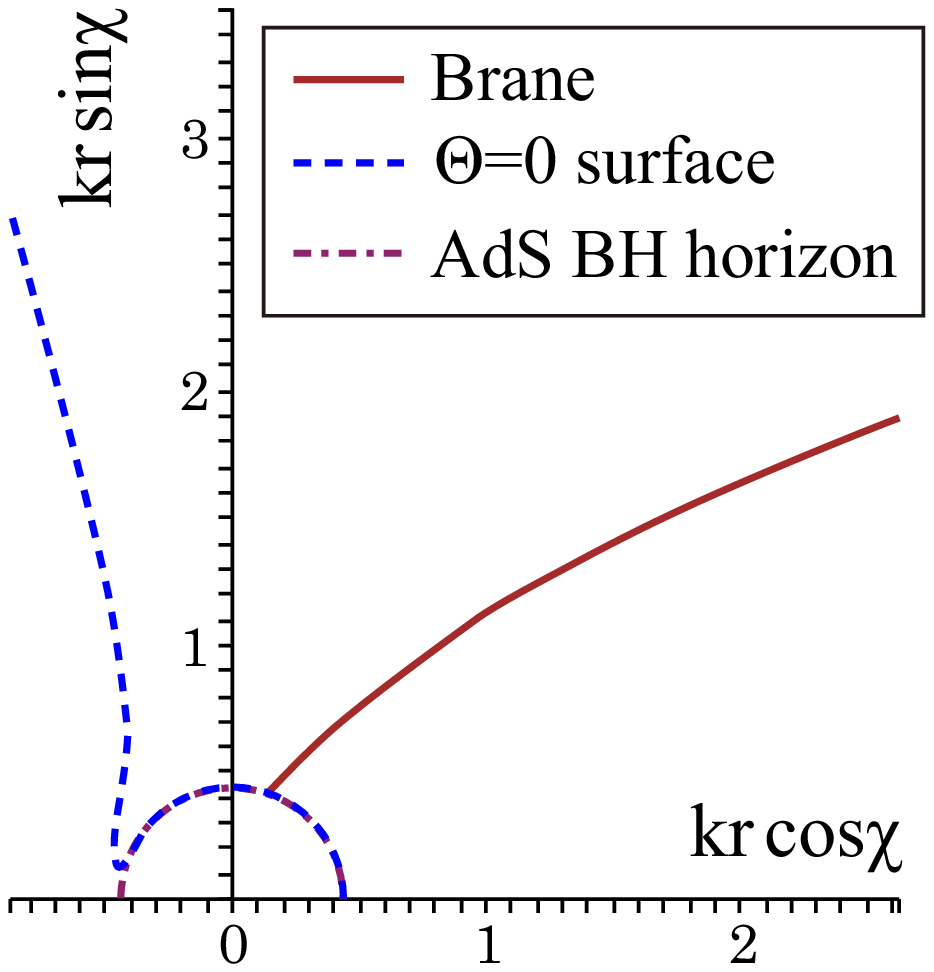}
}
\subfigure[$A_\text{AH}=10^2k^{-3}$]
{\includegraphics[width=5cm, clip]{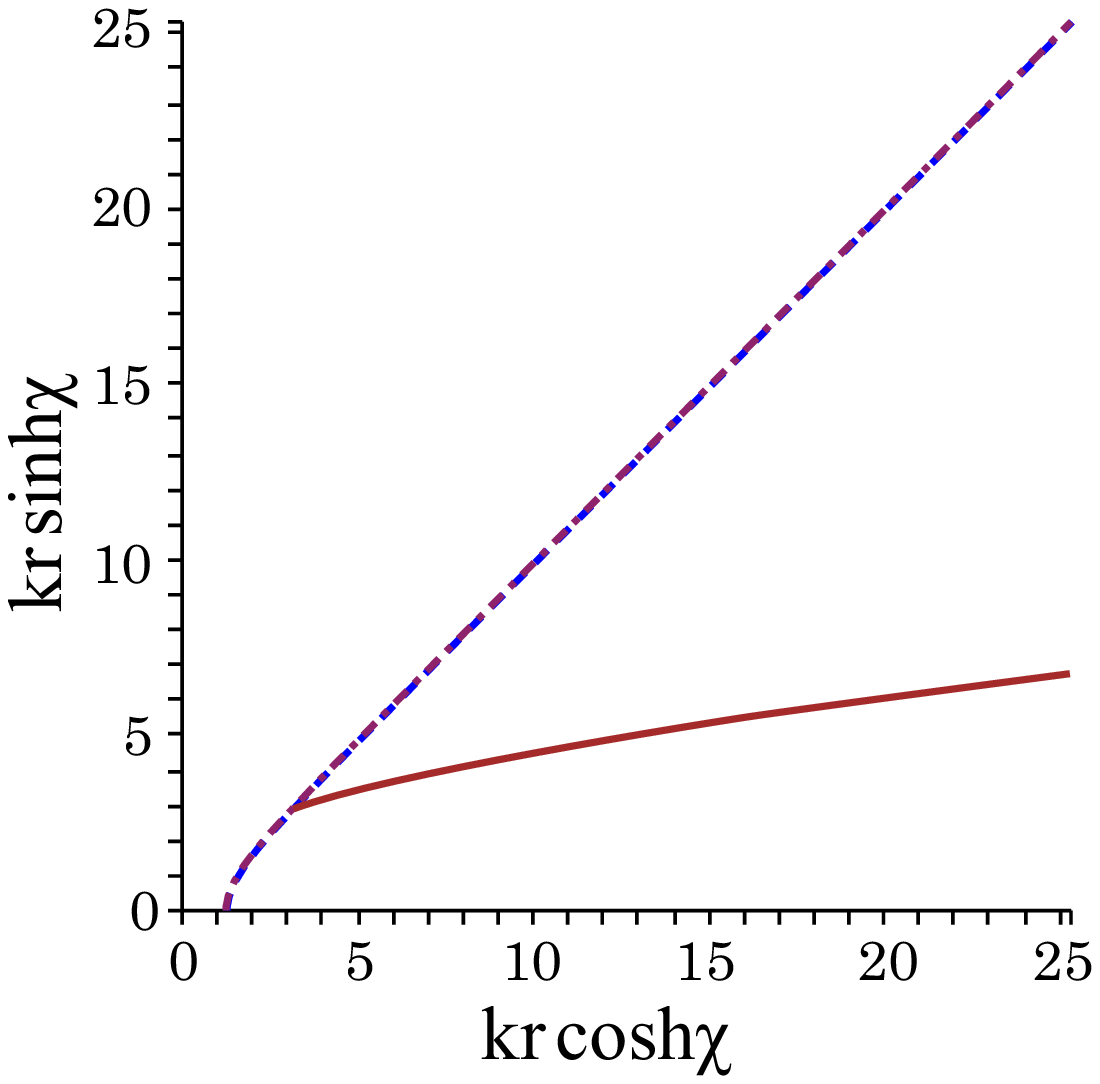}
}
\subfigure[$A_\text{AH}=10^3k^{-3}$]
{\includegraphics[width=5cm, clip]{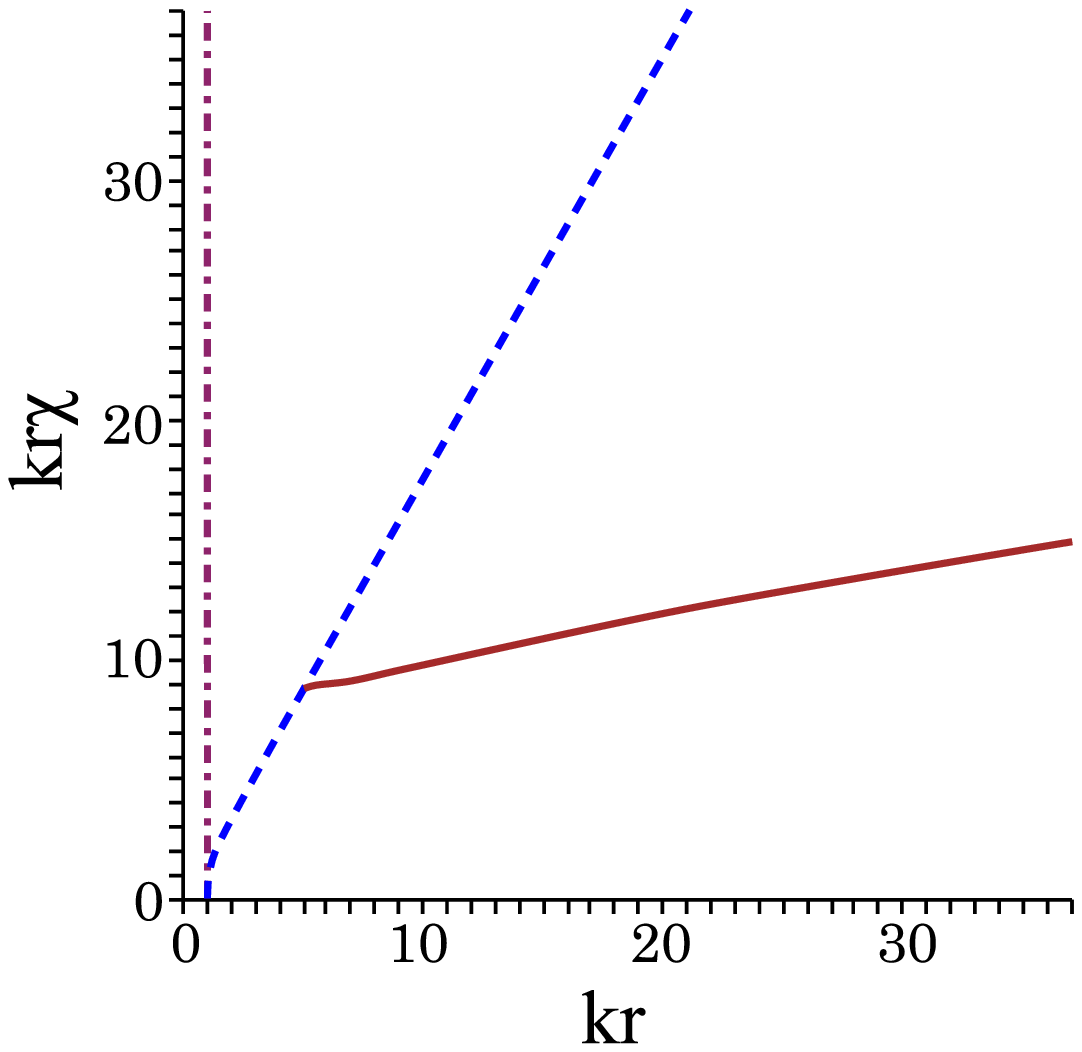}
}
\caption{
The configurations of the AH and the brane in the AdS BH bulk 
for $A_\text{AH}=k^{-3}, 10^2k^{-3}$ and $10^3k^{-3}$. 
The event horizon of the AdS BH bulk is also drawn. 
The type of the AdS BH bulk is different in each plot, i.e.  
spherical, hyperbolic and flat, respectively, in the increasing order of
 $A_\text{AH}$. 
}
\label{rawfig1e}
\end{figure}

\end{widetext}

We found that the mass minimum is realized on the spherical AdS BH 
side when $A_\text{AH}$ is smaller than a typical 
scale determined by the bulk curvature scale, i.e. 
when $A_\text{AH}\lesssim k^{-3}$. 
The minimum seemed to be realized in the limit 
$r_0^\text{AH}\to r_g$ 
in this regime. %(See Appendix.) 
As $A_\text{AH}$ 
is increased, the location of 
the minimum moves toward the hyperbolic AdS BH side.
%,   reaching the point 
%%$(\mu, r^\text{AH}_0)\approx (1,r_g)$. 
%{\color{red} $(\mu, r^\text{AH}_0)\approx (0,r_g)$. }
When $r^\text{AH}_0\approx r_g$, 
AH is very close to the event horizon of the AdS BH bulk. 
If we further increase $A_\text{AH}$, $r^\text{AH}_0$ at 
the mass minimum leaves $r_g$ and the minimum 
is realized in the limit $\mu \to \infty$,
in which the bulk becomes the flat AdS BH spacetime.
%moves to the point
%$(\mu, r^\text{AH}_0)\approx (\infty,(1+\alpha)r_g)$, 
%where $\alpha\ll\mathcal{O}(1)$ is a constant which seems to depend on $A_\text{AH}$.
%For example $\alpha \approx 10^{-2}$ for $A_\text{AH}=10^2$. 
%In this limit $\mu\to \infty$ the bulk becomes
%In this case the bulk becomes the flat AdS BH spacetime. 
In this regime  
the AH deviates from the event horizon of the AdS BH bulk. 
We displayed the trajectories of the AH and the brane 
for the initial data with the minimum ADM mass 
for three typical values of $A_\text{AH}$ in Fig.~\ref{rawfig1e}. 

When we consider a brane-localized BH which is small compared to the bulk 
curvature scale $k^{-1}$, 
the effect of brane tension is negligible. 
Then we have an initial data which has the brane placed on the equatorial 
plane of the AdS BH is approximately a static solution, 
at least, near the horizon. 
Thus such an initial data will have the minimum ADM mass 
among the configurations of the same AH area. 
Our result completely agrees with this expectation. 

\begin{figure}[htbp]
\centering
\includegraphics[width=6cm, clip]{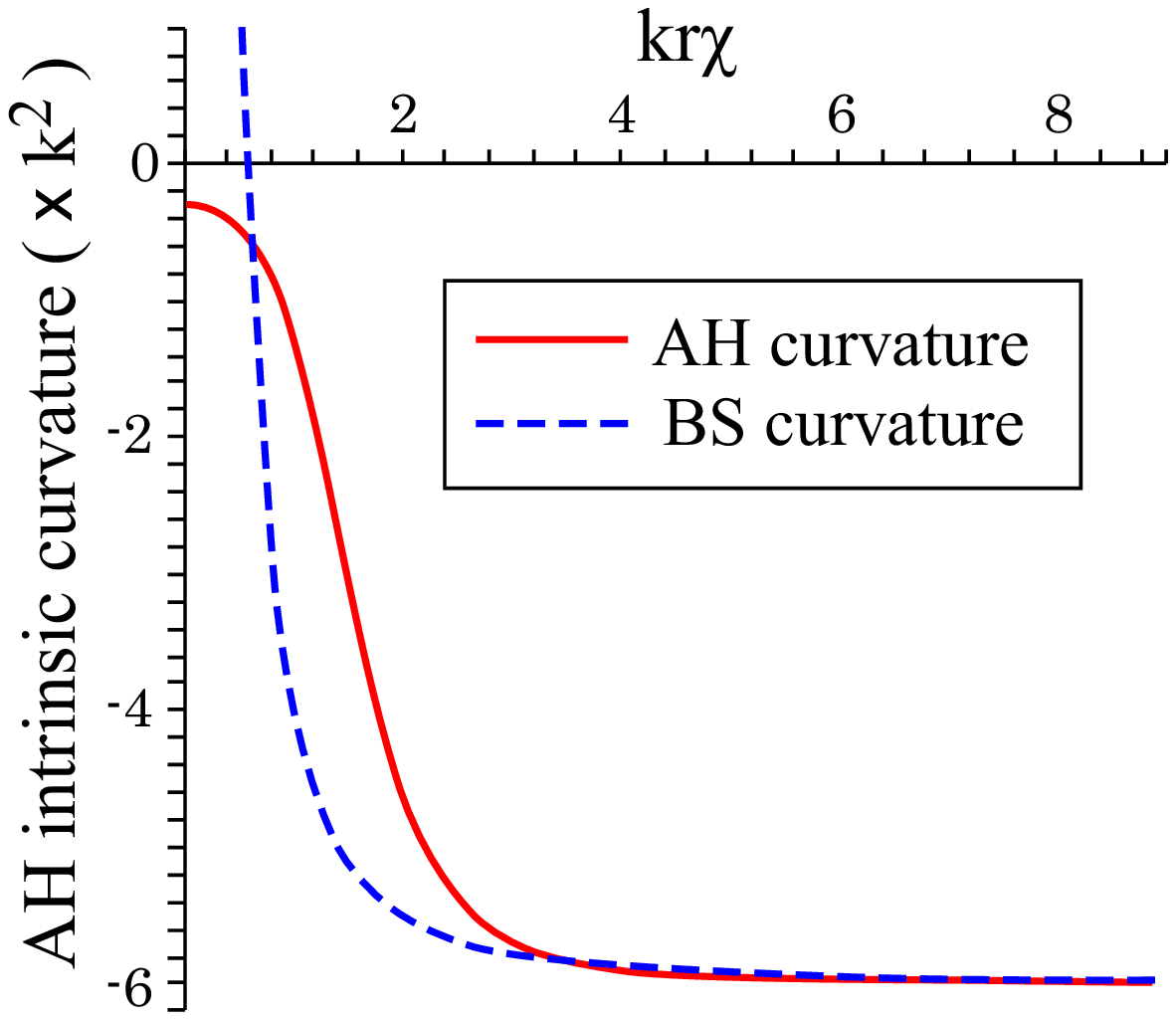}
\caption{
The intrinsic curvature of the induced three metric on the 
AH for the initial data shown in Fig.~\ref{rawfig1e}(c).
The dashed line shows the value of the intrinsic curvature of the BS horizon
whose area is equal to $10^3k^{-3}$.}
\label{rawfig_intAH}
\end{figure}

When the AH is large, the bulk curvature 
effect becomes large. 
Then, five-dimensional AdS BH with the brane on its equatorial plane 
is no longer close to a static configuration. 
In order to study the character of the initial data that 
realizes the minimum ADM mass when $A_\text{AH}\gg k^{-3}$, 
we computed the scalar curvature ${}^{(3)}R_\text{AH}$ 
of the induced three metric on the AH.
We plot ${}^{(3)}R_\text{AH}$ against 
%$\rho=r_AH \chi$, 
%{\color{red} $\rho=r_{AH} \chi$, }
$\rho_\text{AH}$, 
the circumferential radius of $(\theta,\phi)$-two sphere, (normalized 
by the bulk curvature length) in Fig.~\ref{rawfig_intAH}. 
We found that ${}^{(3)}R_\text{AH}$ is negative everywhere on the AH
and the value stays very close to $-6k^2$ near the brane. 
Significant deviation from $-6k^2$ occurs only for 
a small circumferential radius ($\rho\lesssim k^{-1}$).

We compare the above result with 
the intrinsic curvature of the three metric on the BS horizon, 
\begin{equation}
 {}^{(3)}R_\text{BS}=2\left({1\over \rho^2}-3k^2\right). 
\end{equation}
When $\rho\gtrsim k^{-1}$, we have ${}^{(3)}R_\text{BS}\approx -6k^2$. 
This behavior is very similar to the initial data that minimizes 
the mass. 
Thus we find that the AH geometry of our initial data 
is close to the horizon geometry of the BS solution 
near the brane. 

Since BS is an exact static solution, it will give a local or the global 
minimum of the ADM mass for a given AH area. 
Therefore there will be a dip with respect to the mass 
around the BS configuration 
in the complete parameter space of all possible time-symmetric initial data. 
However, we are not here investigating all possible 
configurations of time-symmetric 
initial data. Our parameter space is very restricted. 
In fact, the BS configuration is not contained in our initial data. 
Then, in the regime $A_\text{AH}\gtrsim k^{-3}$, 
the minimum of the ADM mass in our present 
two-dimensional parameter space may correspond to 
the dip around the BS configuration.   
This is a scenario consistent with the conjecture 
that there is no static large brane-localized BH. 
However, it is also possible that the mass minimum that 
we found corresponds to 
a dip associated with another  
static brane-localized BH solution. 

It will be interesting to point out that we can construct 
initial data with AH starting with the pure AdS bulk. 
This corresponds to the case of $\mu=0$ 
on the hyperbolic AdS BH side. This case is not in the 
excluded region of the parameter space. 
(On the spherical AdS BH side, 
$r_g$ behaves like $\approx \sqrt{\mu}$ 
in the limit $\mu\to 0$. 
Thus, in Fig.~\ref{contour5D}  
the location of the starting point of the AHC, $r^\text{AH}_0$, 
is always close to the horizon in this limit.  
Therefore $\mu=0$ does not simply 
correspond to the case of the pure AdS bulk. 
On the other hand, we have $r_g \approx k^{-1}$ in the limit $\mu\to 0$ 
on the hyperbolic AdS BH side. 
Thus, the effect of non-zero $\mu$ on the AHC vanishes in the limit $\mu\to 0$.) 
If we solve Eq.~(\ref{eqAH_}) for $\mu=0$, the AHC becomes 
homogeneous hyperbolic space, as shown in Fig. \ref{rawfig1e}, with 
intrinsic curvature being ${}^{(3)}R_\text{AH}=-6k^2$. 
Its behavior is close to BS for $\rho \gtrsim k^{-1}$,  
but the deviation from BS for $\rho \lesssim k^{-1}$ is larger 
compared with the initial data which minimizes the ADM mass.

\section{Four-dimensional case}
\label{Sec fourD}
In this section 
we analyze the brane-localized BH initial data in four-dimensional
RS-II model in a parallel manner. 
The bulk cosmological constant is $\Lambda=-3k^2$
and the brane tension is $\lambda = 2k/4\pi G_4$,
where $G_4$ is the four-dimensional gravitational constant.
In the four-dimensional case the situation looks quite different 
from the higher-dimensional models. 
The counterpart of the analytic BS solution discussed in 
Sec.~\ref{sec:BS} does not exist. 
Instead, 
there exists an exact solution of a brane-localized BH~\cite{emp2, Emparan:1999fd}. 
Therefore the properties of the initial data 
might be different from the five-dimensional case.

As the bulk spacetime, we consider the four-dimensional AdS BH solution 
\begin{eqnarray}
ds^2=-U(r)dt^2+
%\UIV(r)^{-1}dr^2
\frac{dr^2}{U(r)}
+r^2\tilde{\sigma}_{IJ}(x)dx^Idx^J,
\label{top_metric_4d}
\end{eqnarray}
where 
\begin{eqnarray*}
U(r)=\beta+k^2r^2-\frac{\mu}{r},
\end{eqnarray*}
and 
\begin{eqnarray*}
\tilde{\sigma}_{ij}(x)dx^idx^j=
\begin{cases}
d\chi^2 + \sin^2\chi d\theta^2
& (\beta=+1)~, \\
d\chi^2 + \chi^2 d\theta^2
& (\beta=0)~,
\\
d\chi^2 + \sinh^2\chi d\theta^2
& (\beta=-1)~.
\end{cases}
\end{eqnarray*}
The properties of this solution is similar to the five-dimensional case.
The rescaling corresponding to Eqs.~(\ref{rescaling}) in the present
case is given by 
\begin{equation}
\bar{t}=\left(k\mu\right)^{1/3}t,\quad 
\bar{r}=\left(k\mu\right)^{-1/3}r,\quad
\bar{\chi}=\left(k\mu\right)^{1/3}\chi. 
\end{equation}
The equation that determines the AHC (\ref{eqAH}) becomes 
\begin{align}
 \frac{\sqrt{U}r_\text{AH}}{r'_\text{AH}} \chi''_\text{AH}
+&3\sqrt{U}\chi'_\text{AH}
-\frac{\cot\chi_\text{AH}}{\sqrt{U}r_\text{AH}}r'_\text{AH}
=0, \notag \\
-\frac{1}{\sqrt{U}r_\text{AH}\chi'_\text{AH}} r''_\text{AH}
&+ 2\sqrt{U}\chi'_\text{AH}
\notag \\
+\frac{r'_\text{AH}}{2U^{3/2}r_\text{AH}\chi'_\text{AH}}
&\left( 
%U'
\frac{dU}{dr}r'_\text{AH}
-2 U\chi'_\text{AH} \cot\chi_\text{AH} \right)
=0.
\label{eqAH4D}
\end{align}
The Hamiltonian constraint on the brane~(\ref{HC})
is given by Eq.~(\ref{eqAH4D}), replacing 
$r_\text{AH}(\zeta)$ and $\chi_\text{AH}(\zeta)$ with $r_b(\xi)$ and $\chi_b(\xi)$
on the left hand sides,
and $0$ with $-2k$ on the right hand sides.
The area of the AH is given by 
\begin{equation}
{}^{(4)}A_\text{AH}
%= 2\int_0^{\chi^b_0} 2\pi 
%\sqrt{r_\text{AH}^2+\UIV^{-1} r'_\text{AH}\!\!\!\!\!^2~\,}
%\;r_\text{AH} \sin \chi \; d\chi.
= 2\int_0^{\zeta_0} 2\pi 
%r_\text{AH}\sin\chi_\text{AH} \; 
\rho_\text{AH}
\,d\zeta.
\label{4DAHarea}
\end{equation}
%where $\rho_\text{AH}(\zeta)$ is defined in the same way with the five-dimensional case.

On the three-dimensional brane, the mass of an object is 
given by 
\begin{equation}
M_3= \frac{\delta\phi}{4\pi k G_4},
\label{defM3}
\end{equation}
where $\delta \phi$ is the deficit angle 
in the asymptotic region on the brane. 
Notice that $kG_4/2$ is the three-dimensional effective 
Newton constant on the brane.
The asymptotic deficit angle $\delta\phi$ is 
obtained from the relationship
between the proper length in the radial direction 
%$\tilde{l}_{rad}=\int d\xi$, 
$\xi$,
and the circumferential radius 
%$\tilde{l}_\text{circ}$ 
$\rho_b$ as
\begin{equation}
%    d \tilde{l}_\text{circ} = (2\pi - \delta \phi)d \tilde{l}_{rad}.
    d \rho_b = (2\pi - \delta \phi)d \xi.
\end{equation}
Thus, the explicit formula for $M_3$ is
\begin{equation}
M_3 = {1\over 2 k G_4} 
\lim_{\xi\to \infty}\left(1-\rho_b'\right).
\end{equation}

The analysis of the initial data goes in parallel to 
the five-dimensional case presented in Sec.~\ref{results}.
In the five-dimensional case we used the BS solution as a 
reference. As mentioned above, 
in the four-dimensional RS-II brane world a similar 
analytic BS solution does not exist, but 
there is an exact solution of the 
brane-localized BH found by 
Emparan, Horowitz and Myers~\cite{emp2} (EHM). 
Here we briefly review the property of this exact solution. 

The bulk of the solution is given by a special case of 
the AdS C-metric~\cite{Plebanski:1976gy}
\begin{eqnarray}
&&    ds^2=\frac{1}{k^2(x-y)^2}
\left[
-F(y)dt^2
+\frac{dy^2}{F(y)}\right.\cr
%- \left(   y^2+2\bar{\mu} y^3 \right) dt^2
%+ \frac{dy^2}{\left( y^2+2\bar{\mu} y^3 \right)}
&&\qquad\qquad \qquad\qquad \qquad 
\left.+ \frac{dx^2}{G(x)}
+ G(x)d\phi^2 \right],\qquad 
\label{C-metric}
\end{eqnarray}
where $F(y)=y^2+2\bar{\mu} y^3$ and
$G(x)=1-x^2-2\bar{\mu} x^3$.
This metric has a BH horizon at 
$y=y_\text{hor}\equiv -1/2\bar{\mu}$.
Basically the coordinate $y$ 
is the radial coordinate, 
and $x$ is an angular coordinate.
In the case of $\bar{\mu}>0$, $G(x)=0$ has   
three roots. We denote the largest positive root 
by $x_2$. We take the region specified by 
$0\leq x\leq x_2$ and $-1/2\mu\leq y\leq x$ 
as the bulk. The $\mathbb{Z}_2$-symmetric brane is 
located at $x=0$. Then the junction condition across 
the brane is satisfied. 
The axis of $\phi$-rotation is at $x=x_2$.  
To avoid the conical singularity on this axis,  
the periodicity of $\phi$ should be chosen as 
\begin{equation}
 \Delta \phi = \frac{4\pi}{\left|G'(x_2)\right|}
=\frac{2\pi}{x_2\left(1+3\bar{\mu} x_2\right)}~.
\label{period}
\end{equation}
Thus, the horizon area of the EHM solution is given by 
\begin{align}
A_\text{EHM}
&=2\Delta \phi \int_0^{x_2}\frac{dx}{k^2(x-y_\text{hor})^2}
\notag \\
&=\frac{16\pi \bar{\mu}^2 }{k^2(1+3\bar{\mu} x_2)(1+2\bar{\mu} x_2)}.
\label{areaEHM}
\end{align}
It is good to consider the limiting cases, 
$\bar{\mu} \ll 1$ and $\bar{\mu} \gg 1$, to understand 
the behavior of $A_\text{EHM}$. 
When $\bar{\mu}\ll 1$, $x_2$ and $A_\text{EHM}$ are approximated as
\begin{equation}
 x_2\simeq 1-\bar{\mu}, \qquad
A_\text{EHM}\simeq 16\pi\bar{\mu}^2k^{-2}.
\end{equation}
When $\bar{\mu}\gg 1$, 
$x_2$ and $A_\text{AH}$ are approximately given by 
\begin{equation}
 x_2\simeq \left(2\bar\mu\right)^{-1/3}, \qquad
A_\text{EHM}\simeq \frac{8\pi}{3k^2}\left(2\bar \mu\right)^{2/3}.
\end{equation}

The mass $M_3$ 
measured on the brane is proportional 
to the asymptotic deficit angle $\delta\phi$ as given in 
Eq.~(\ref{defM3}). 
Since the induced metric on the brane becomes 
\begin{equation}
 d\tilde s^2=-\left(1+{2\bar\mu\over r}\right)d\bar t^2
    +{1\over 1+{2\bar\mu\over r}} dr^2 +r^2 d\phi^2, 
 \end{equation}
where we introduced $\bar t\equiv k^{-1}t$ and $r\equiv 1/ky$,  
we find that the deficit angle is simply given by 
$\delta\phi=2\pi -\Delta\phi$.     
Hence, we have 
\begin{equation}
 M_3={1\over 2k G_4}(1-F), 
\label{MEHM}
\end{equation}
with $F\equiv\Delta\phi/2\pi=x_2^{-1}(1+3\bar\mu x_2)^{-1}$. 
Using $F=1-2kG_4 M_3$, 
the area of the EHM solution (\ref{areaEHM}) can be expressed as
\begin{equation}
A_\text{EHM}=\frac{16\pi}{9k^2}
\frac{\left(-2+\sqrt{1+3F^2}\right)^2}{-1+\sqrt{1+3F^2}}.
\label{AEHM}
\end{equation}
We can easily relate the mass and the horizon area of 
the EHM BH solution by using the formulas (\ref{MEHM})
 and (\ref{AEHM}). 

We can also calculate the intrinsic curvature of the two-dimensional
metric on the event horizon as 
\begin{align}
{}^{(2)}R_\text{EHM}=
 \frac{k^2}{2\bar\mu^2}
&\Bigl\{
8\left(\bar\mu x\right)^3
+12\left(\bar\mu x\right)^2
\notag \\
&\qquad\;\;
+6\left(\bar\mu x\right) - 4\bar\mu^2+1
\Bigr\}, 
\label{EHM intrinsic}
\end{align}
and $x$ is related to the radius of the circle in 
the $\phi$-direction, $\rho$, as 
\begin{eqnarray*}
 \rho={\sqrt{G(x)}\over k(x-y_\text{hor})}{\Delta\phi\over 2\pi}
   ={2\bar\mu\over k}{\sqrt{1-x^2-2\bar\mu x^2}\over 
        x_2(1+3\bar\mu x_2)(1+2\bar\mu x)}.
\end{eqnarray*}

Figure \ref{Fig area4D} shows the parameter region in which the 
AH area becomes larger than 
the horizon area of the EHM solution with the same mass.
We found that the whole  parameter region in which the AH area becomes larger than
the horizon area of the EHM solution is in the excluded region. 
This result is consistent with the naive expectation that 
the EHM solution is the most stable black object in the four-dimensional 
RS-II model and hence that any other black objects, even if they exist, 
have smaller horizon size.

In Fig.~\ref{contour4D}
we show the plots of the mass 
normalized by the mass of the EHM solution of the same horizon area, which are 
parallel to Fig.~\ref{contour5D}
of the five-dimensional case. 
We found that the behavior of the contour plot 
is similar to that of the five-dimensional case.
When we vary the value of the AH area,  
the location of the mass minimum changes exactly in the same way 
as in the previous case. 

We show the configurations of the branes and AHs for the initial data 
of the mass minimum for a few fixed values of $A_\text{AH}$ in 
Fig.~\ref{rawfig_4d}. 
The shape of the brane and the AH is also 
similar to the five-dimensional case.
In Fig.~\ref{rawfig_intAH_4d} we plot the intrinsic curvature 
on the AH for $A_\text{AH}=10^2k^{-2}$ by the solid curve. 
The counterpart in the EHM solution is also shown in the same plot.

The interpretation of these results goes almost parallel to 
the five-dimensional case.
The difference appears when the AH area is large. 
We consider the BS solution as a reference 
in the five-dimensional case, while the EHM solution 
in the four-dimensional case. 
The behavior of these two solutions look quite different 
at first sight. The horizon of the BS solution extends 
infinitely, while that of the EHM solution is finite. 
However, if we compare the intrinsic curvature of the 
horizon of these two exact solutions shown in 
Fig.~\ref{rawfig_intAH} and Fig.~\ref{rawfig_intAH_4d}, 
the behavior is quite similar except for the region 
where the circumferential radius $\rho$ is small compared 
with the bulk curvature length scale $k^{-1}$. 

\begin{widetext}

\begin{figure}[htbp]
\centering
\subfigure[Spherical AdS BH background, $\mu=10^{-2}k^{-1}$
]
{\includegraphics[width=6.4cm, clip]{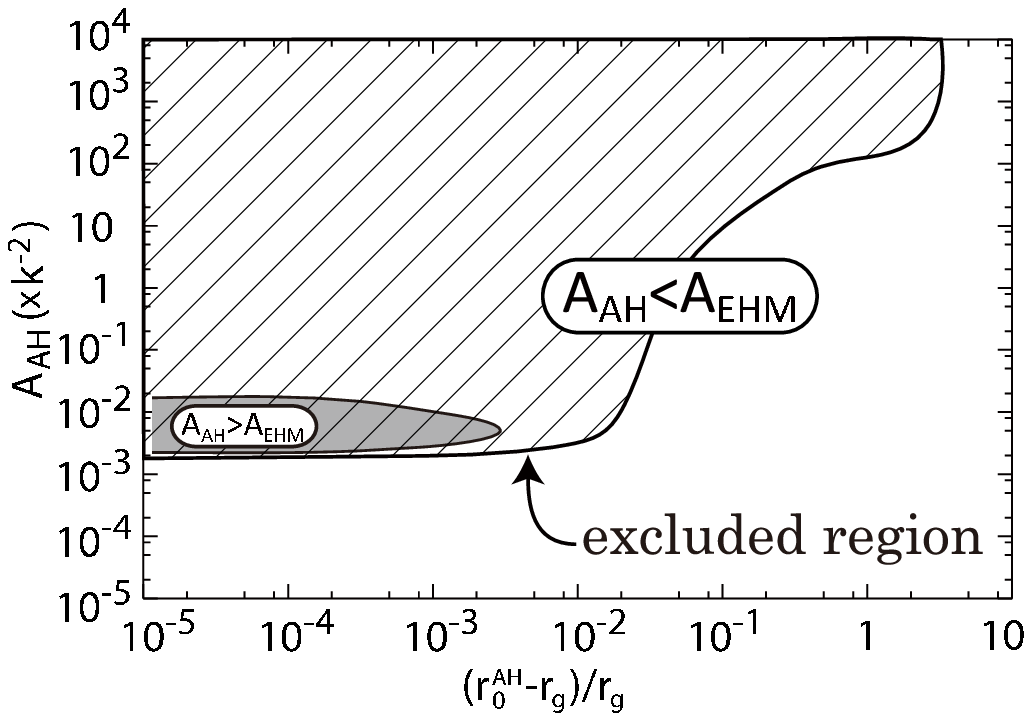}
\label{mu1e-2_4d.eps}}
\subfigure[Spherical AdS BH background, $\mu=10k^{-1}$]
{\includegraphics[width=6.4cm, clip]{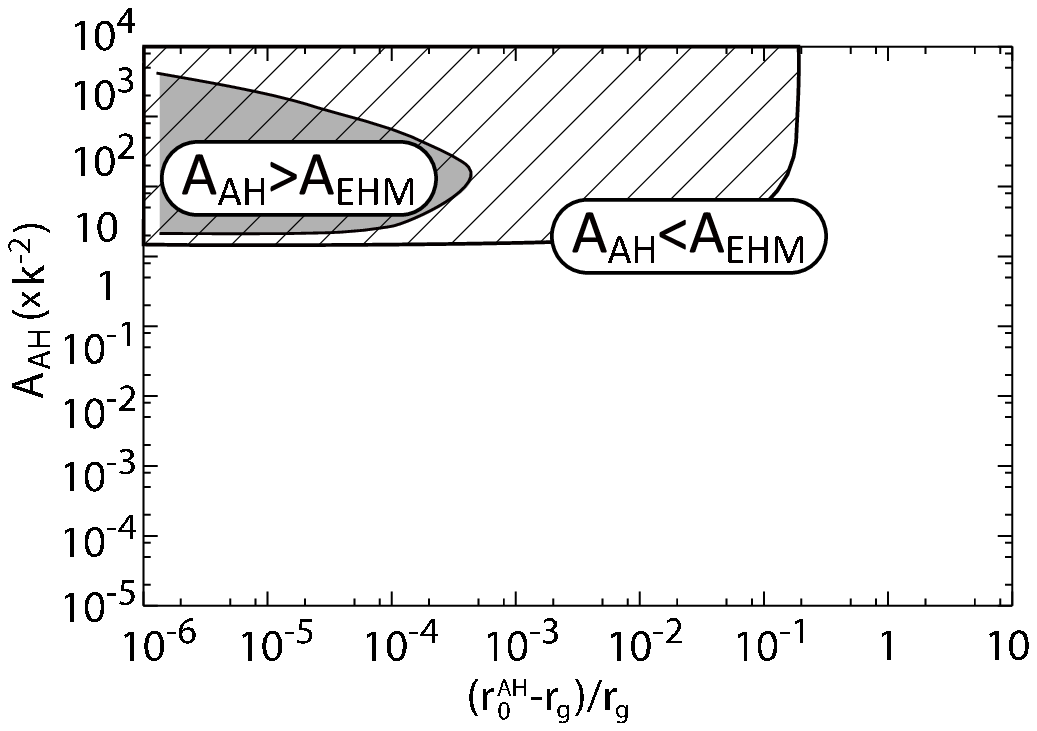}
\label{mu1e1_4d.eps}}
\subfigure[Hyperbolic AdS BH background, $\mu=(-2/3^{3/2}+10^{-2})k^{-1}$]
{\includegraphics[width=6.4cm, clip]{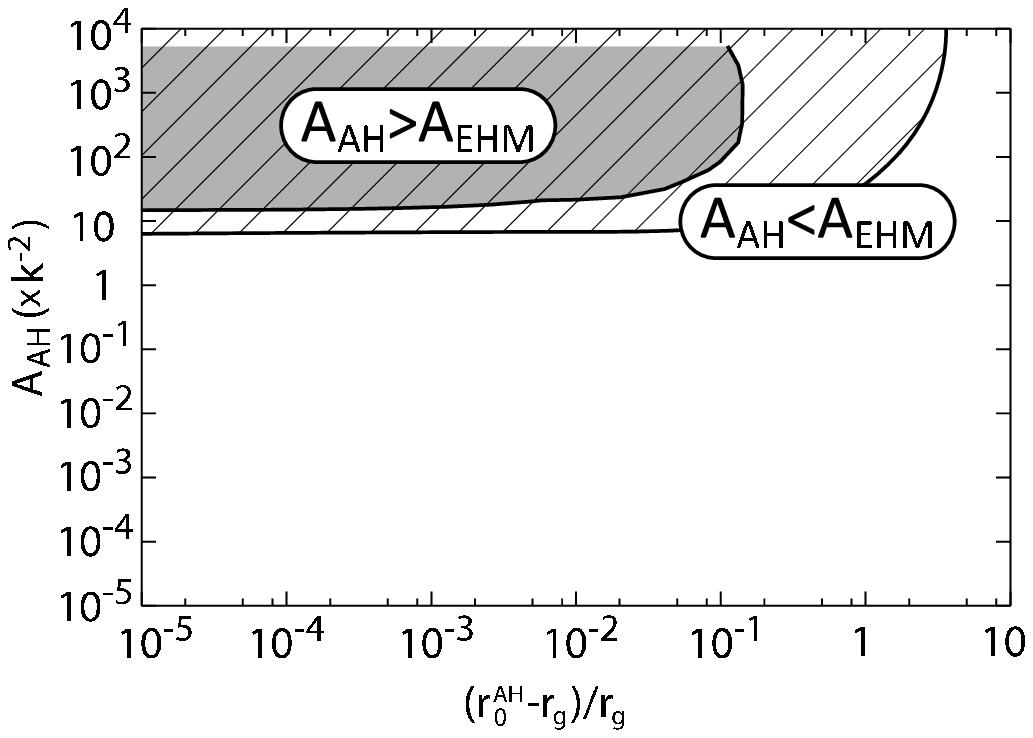}
 \label{mu1e-1_4d.eps}}
\subfigure[Hyperbolic AdS BH background, $\mu=10k^{-1}$]
{\includegraphics[width=6.4cm, clip]{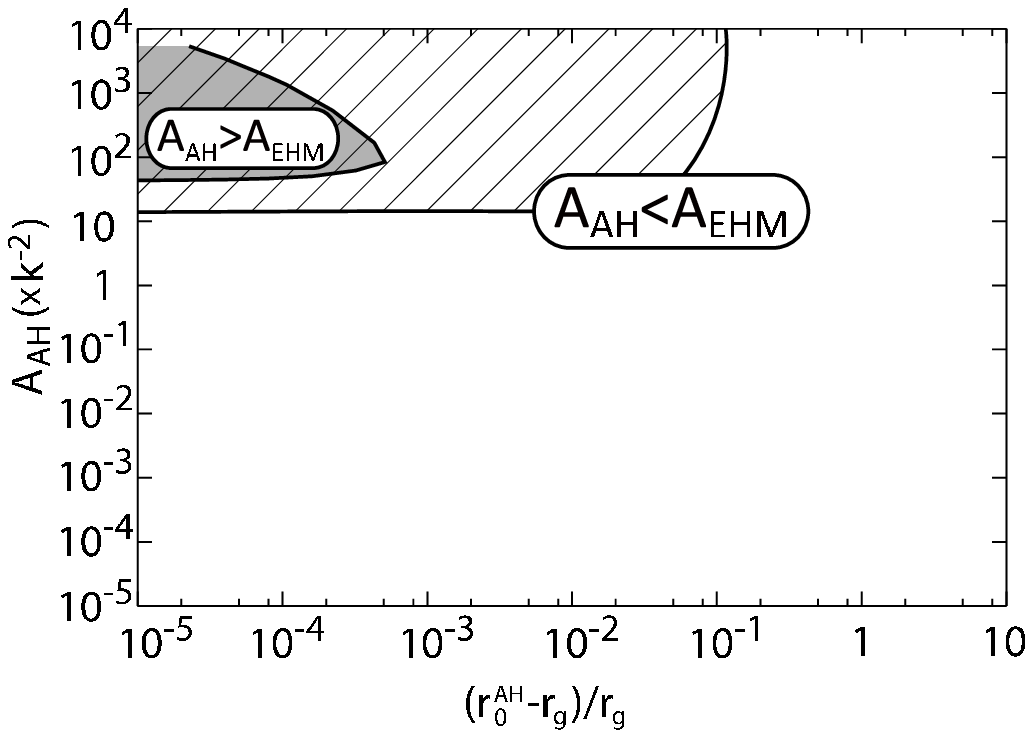}
\label{mu1e0_4d.eps}}
\caption{
The result of area comparison between the AH in the initial data
and the EHM solution for 
%$\mu=10^{-2}, 10^{-1}, 1, 10$.
$\mu=10^{-2}k^{-1}$ and $10k^{-1}$ in the spherical AdS BH background case,
and $\mu=(-2/3^{3/2}+10^{-2})k^{-1}$ and $10k^{-1}$ in the hyperbolic AdS BH background case.
The AH area of the initial data is larger than the horizon area of the EHM solution 
for the parameters in the gray regions in the plots,
and in the other regions the horizon area of the EHM solution is larger.
The shaded regions are excluded from the analysis because
the initial data in these regions have outer AHs.
For any $\mu$, the gray region is covered by a shaded region.
It means that the horizon area of the EHM solution 
is larger than the AH area of the initial data
for any parameters.
}
\label{Fig area4D}
\end{figure}

\begin{figure}[htbp]
\centering
\subfigure[$A_\text{AH}=k^{-2}$ ($M_\text{EHM}\simeq 0.119k^{-1}G_4^{-1}$)
]
{\includegraphics[width=8.3cm, clip]{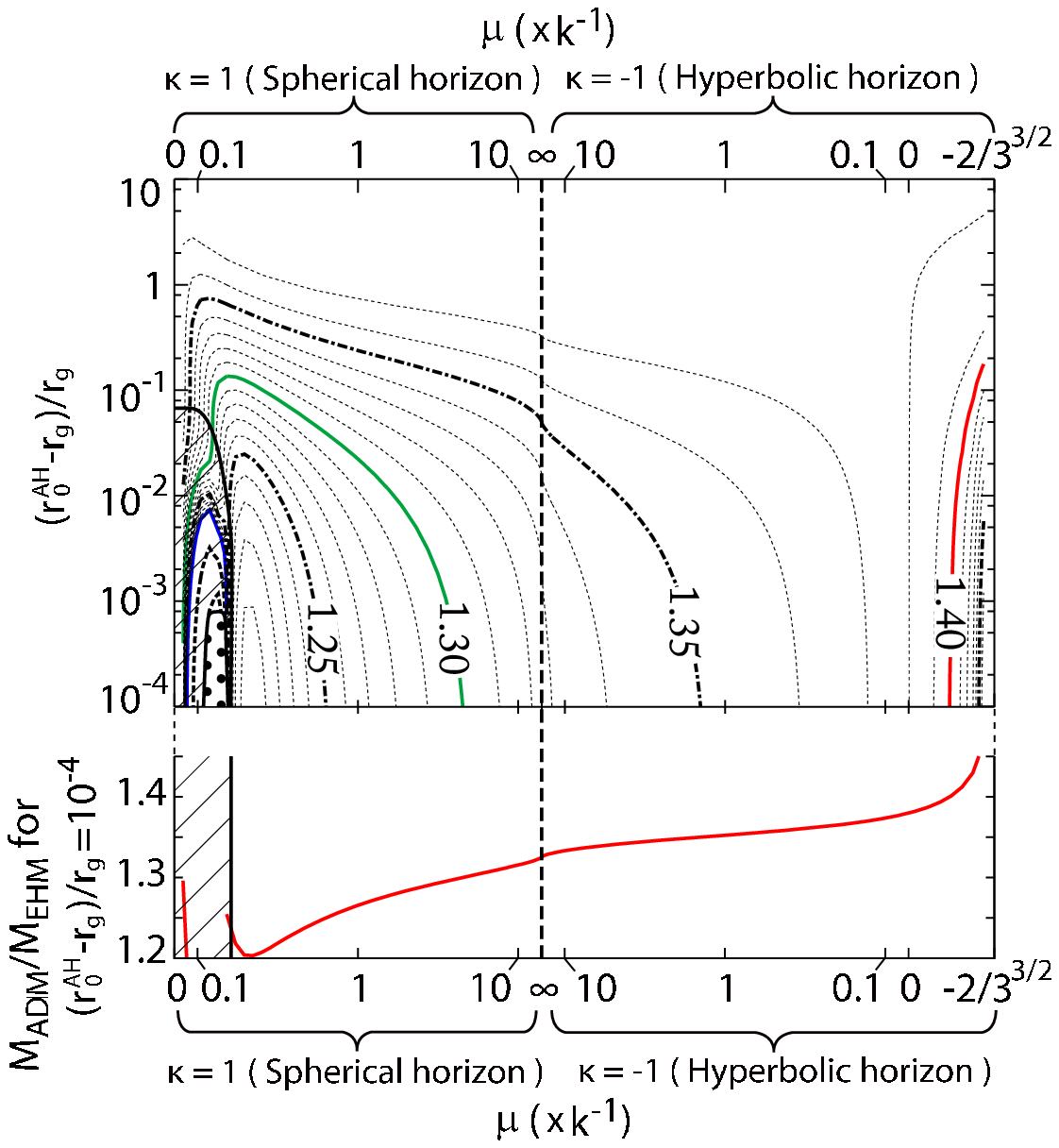}
}
\subfigure[$A_\text{AH}=10k^{-2}$ ($M_\text{EHM}\simeq 0.267k^{-1}G_4^{-1}$)]
{\includegraphics[width=8.3cm, clip]{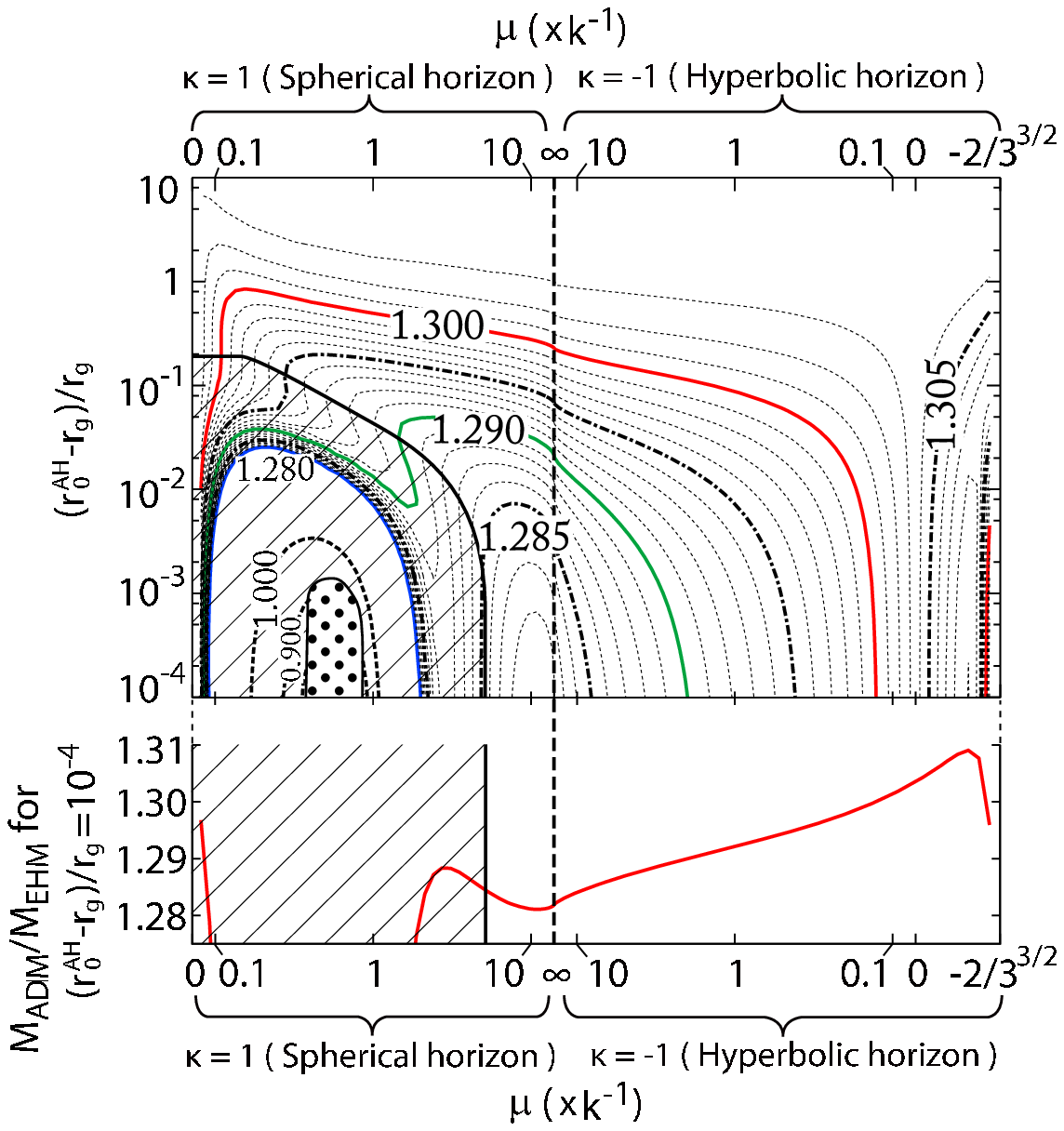}
}
\subfigure[$A_\text{AH}=30k^{-2}$ ($M_\text{EHM}\simeq 0.343k^{-1}G_4^{-1}$)]
{\includegraphics[width=8.3cm, clip]{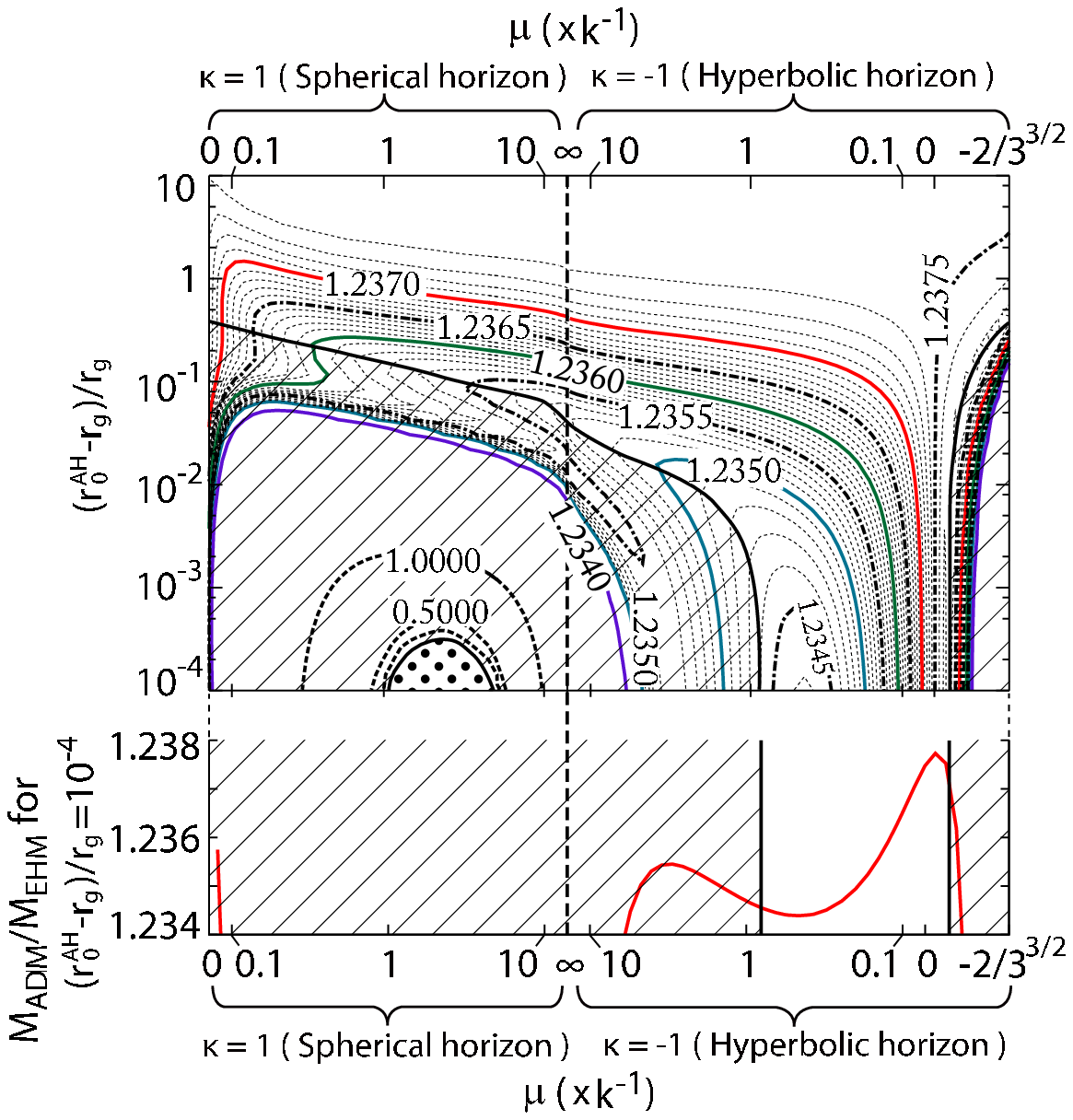}
}
\subfigure[$A_\text{AH}=10^{2}k^{-2}$ ($M_\text{EHM} \simeq 0.407k^{-1}G_4^{-1}$)]
{\includegraphics[width=8.3cm, clip]{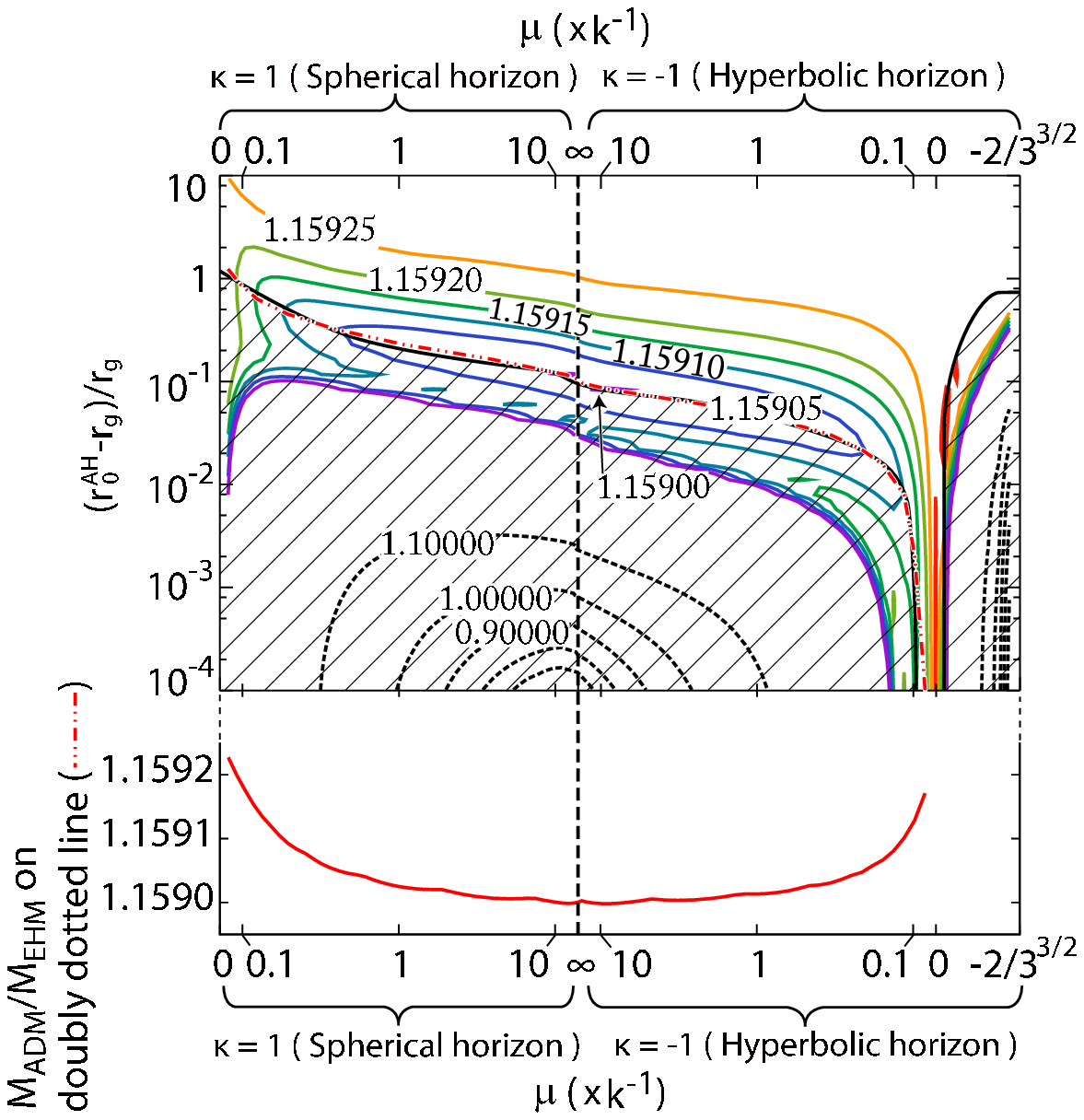}
}
\caption{The contour plots of $M_\text{ADM}/M_\text{EHM}$ 
in the four-dimensional case, where $M_\text{EHM}$ is the mass of the 
EHM BH whose horizon area is equal to $A_\text{AH}$.
The curves below the contour plots in the panels
(a), (b) and (c) show 
the values of the ADM mass for
$(r_0^\text{AH}-r_g)/r_g=10^{-4}$,
and the curve in the panel (d) 
shows the value along the doubly dotted curve in the contour plot.
}
\label{contour4D}
\end{figure}

\begin{figure}[htbp]
\centering
\subfigure[$A_\text{AH}=k^{-2}$]
{\includegraphics[width=5cm, clip]{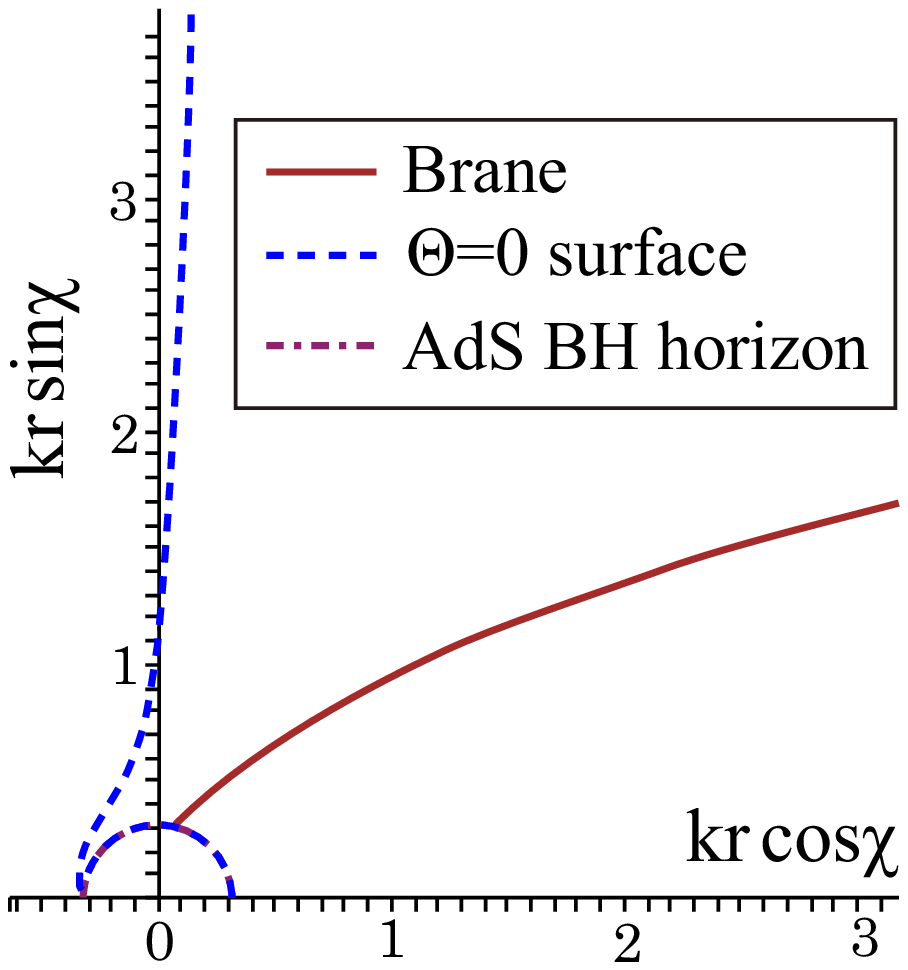}
\label{rawfig1e0_4d}}
\subfigure[$A_\text{AH}=30k^{-2}$]
{\includegraphics[width=5cm, clip]{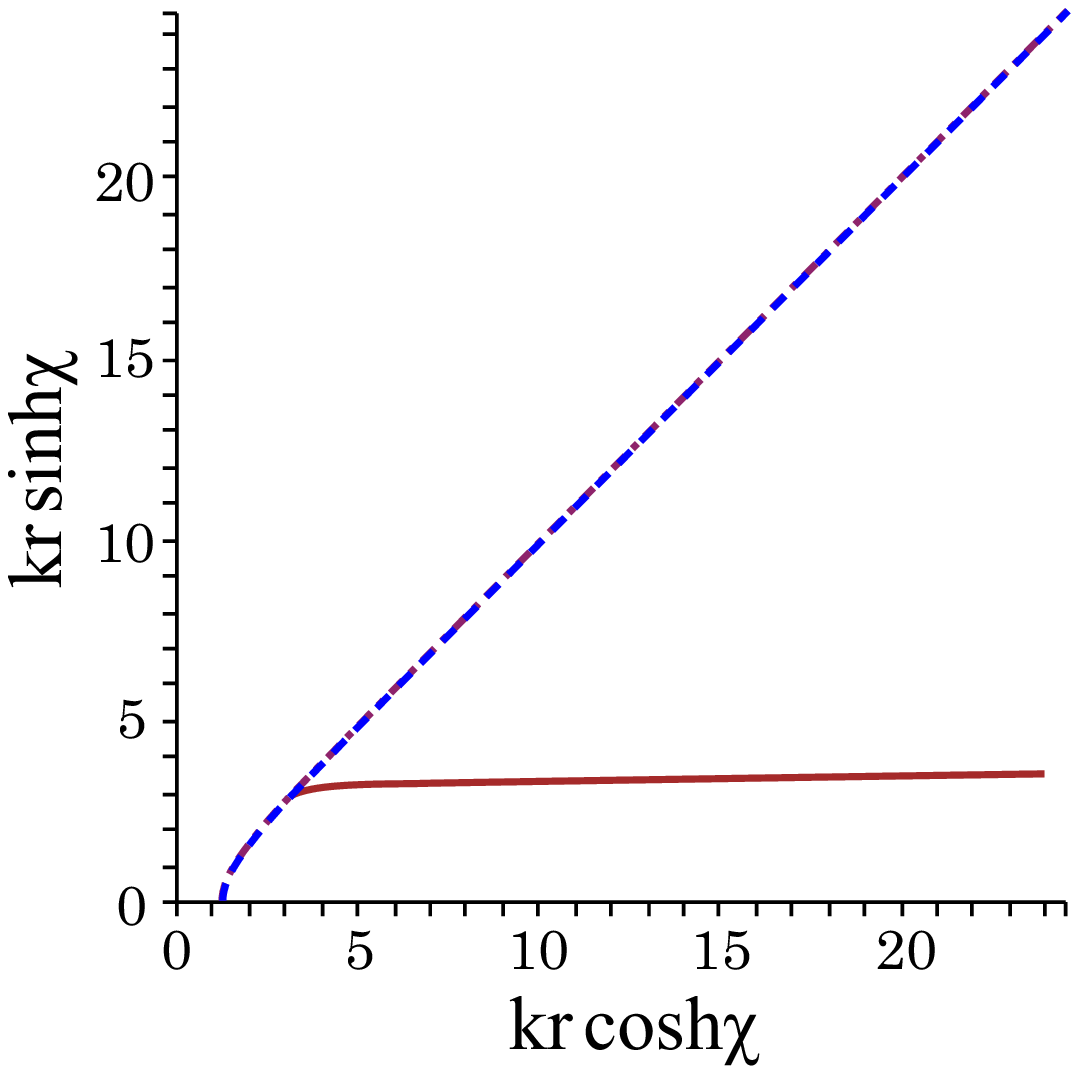}
 \label{rawfig3e1_4d}}
\subfigure[$A_\text{AH}=10^2k^{-2}$]
{\includegraphics[width=5cm, clip]{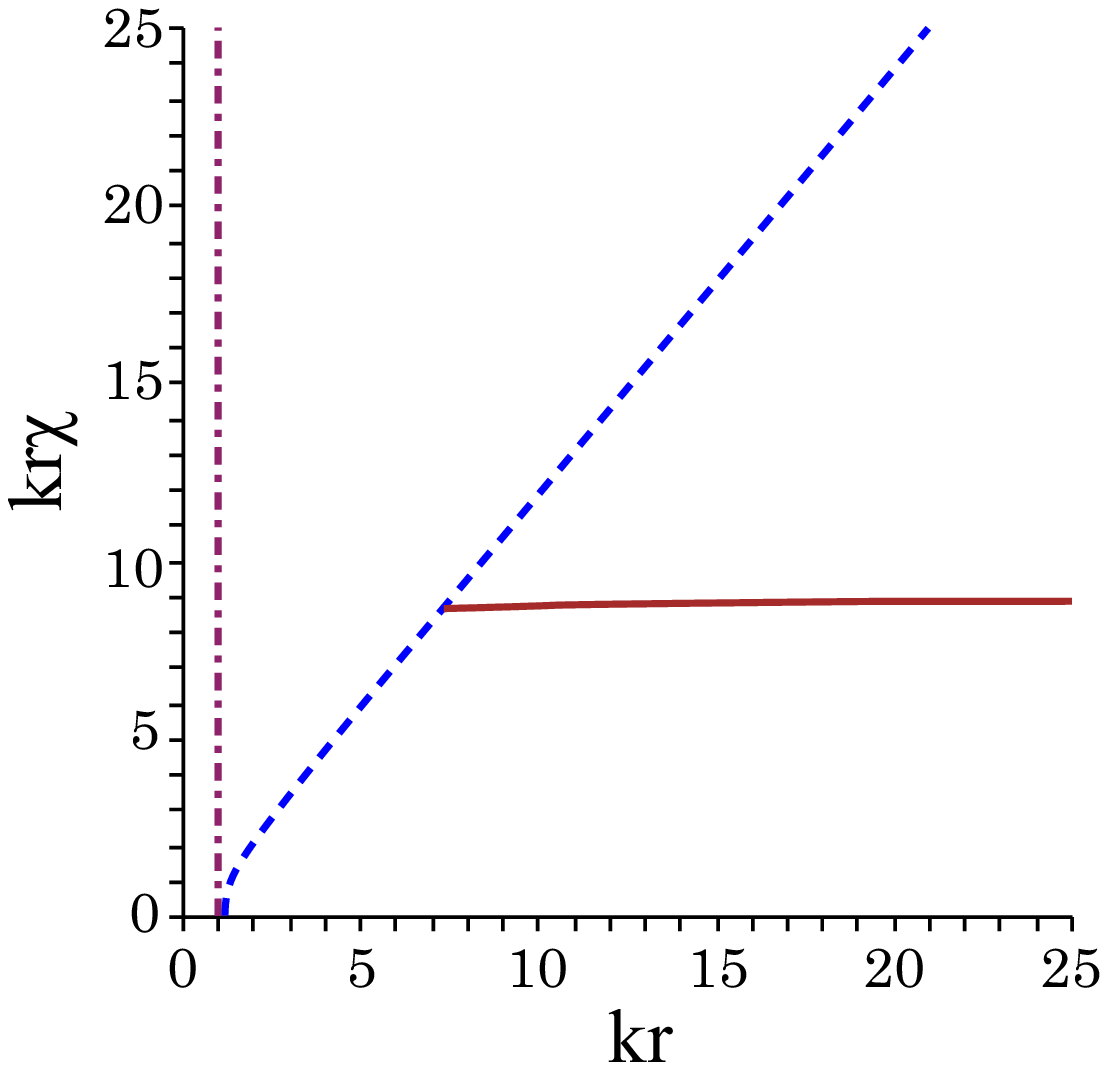}
\label{rawfig1e2_4d}}
\caption{
The configuration of the initial data with the minimum mass 
for a few values of $A_\text{AH}$.
The bulk metrics of these plots are spherical, hyperbolic and flat AdS
 BHs, respectively.
}
\label{rawfig_4d}
\end{figure}

\end{widetext}

\begin{figure}[htbp]
\centering
{\includegraphics[width=6cm, clip]{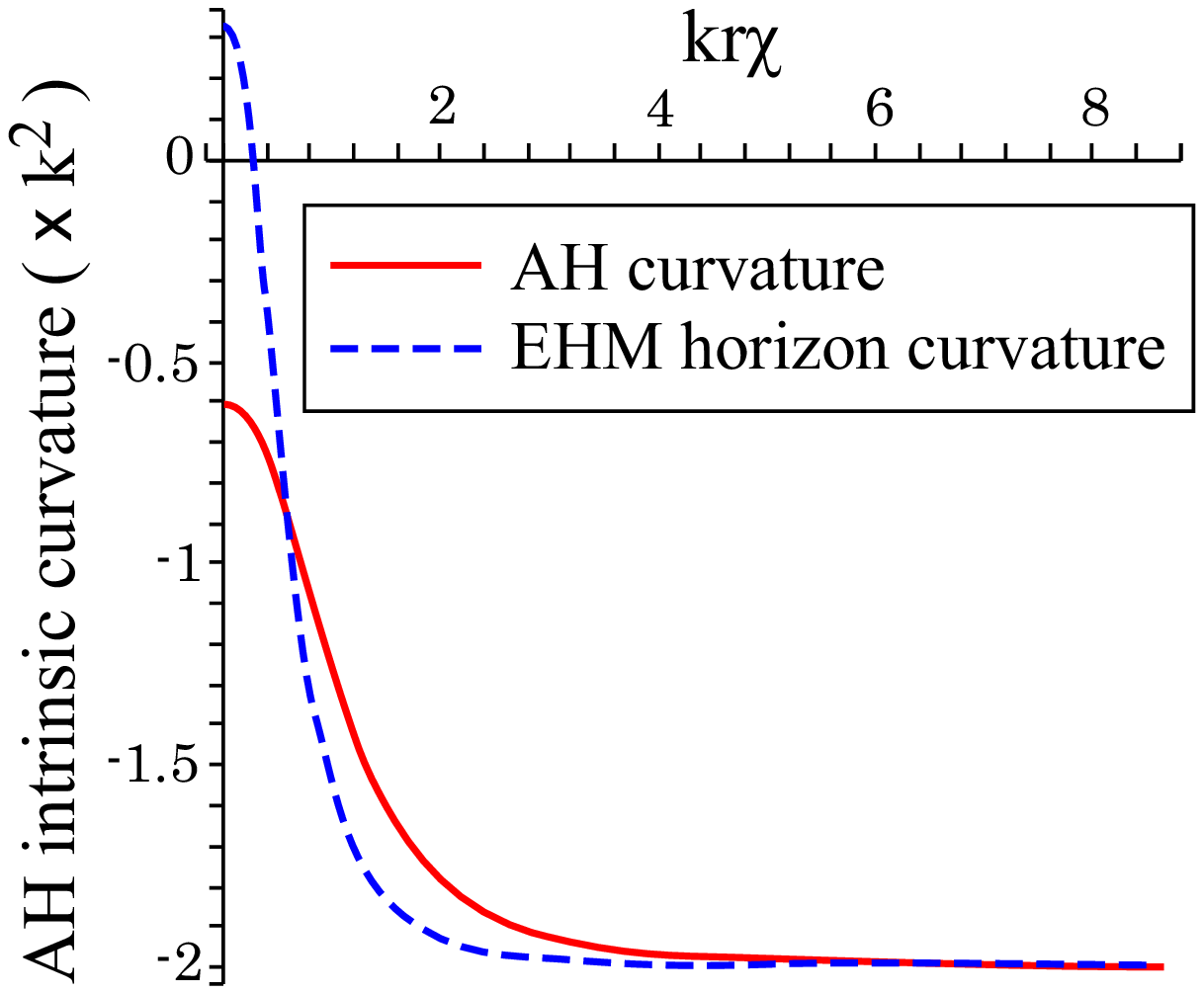}
\caption{ 
${}^{(2)}R_\text{AH}$ of the the AH in the panel (c) in
 Fig.~\ref{rawfig_4d}.
The counterpart in the EHM solution is also shown by the dashed line. 
}
\label{rawfig_intAH_4d}}
\end{figure}

\section{Discussion}

In order to have an insight into the black objects  
with a large horizon radius in the model of warped 
extra-dimension, we studied time-symmetric 
initial data of RS-II model in five and 
four-dimensions.  
We constructed three-parameter family of time-symmetric 
initial data with a brane-localized AH, by placing 
a $\mathbb{Z}_2$-symmetric brane in 
the AdS Schwarzschild bulk. 
We have shown that one can construct 
initial data 
with an arbitrarily large AH area. 
Our method of constructing the initial data requires 
just solving ordinary differential equations. 
Hence, the conclusion that initial data 
with an arbitrarily large AH area exist is quite robust. 

In the five-dimensional model the area of 
the obtained initial data was always smaller than 
the event horizon area of the BS solution whose mass is 
as large as the initial data when the 
horizon area is sufficiently large compared to the 
bulk curvature length cubed.  
In the following sense,
this result is consistent with the scenario that 
these initial data evolve into configurations 
similar to BS, which is unstable through Gregory-Laflamme 
instability. The event horizon, which must exist outside 
the AH area, should have a larger area than AH. 
The area theorem tells that the area of the event horizon 
does not decrease as a course of time evolution, while 
the mass will not increase when there is no incoming 
energy flux from infinity.  
Hence, if there is an initial data whose horizon area  
is significantly larger than that of BS with 
the same mass, such an initial data cannot evolve 
into the configuration close to BS. 

We also explored the initial data that realizes the minimum 
mass with the AH area fixed, which is expected to be the closest
to a static solution. 
We found that the initial data that realizes 
the minimum mass is close to the spherical AdS BH with a brane on its equatorial
plane when the AH area is small, 
while it is close to BS when the AH area is large.

We conducted the same analysis also in the case of 
four-dimensional bulk spacetime since the situation 
looks quite different in this case. 
The BS solution does not exist, but 
we have an
exact brane-localized BH solution found by Emparan, Horowitz and Myers (EHM), instead. 
Nevertheless, the results of the analyses as to the 
time-symmetric initial data were quite similar to 
the five-dimensional case. 
We found that the area of the initial data
is always larger than the EHM solution with the same mass, 
which is in harmony with the naive expectation 
that the EHM solution is the most stable
black object in four-dimensional RS-II model.

Unfortunately, the analysis presented in this paper did not provide
a strong indication about the 
classical evaporation conjecture 
because the initial data that we examined were very limited. 
However, it is a remarkable progress that we have shown 
that time-symmetric initial data with a large AH area 
can be constructed. 
As a next step, we can consider the time evolution
of these initial data. 
The family of initial data we constructed in this study will 
be a good starting point for researches in this direction.

\begin{center}
\bf{Acknowledgments }
\end{center}
TT is supported by Monbukagakusho Grant-in-Aid for
Scientific Research Nos. 17340075 and 19540285. This
work is also supported in part by the 21st Century COE
``Center for Diversity and Universality in Physics'' at Kyoto
university, from the Ministry of Education, Culture,
Sports, Science and Technology of Japan. The authors
thank Takashi Nakamura, 
Shuichiro Yokoyama and Keisuke Izumi for useful comments.

\end{document}